\documentclass[aps,prd,twocolumn,showpacs]{revtex4-1}

\usepackage{wallpaper}
\usepackage[colorlinks,plainpages=false]{hyperref}
\usepackage{amsmath}
\usepackage{upgreek}

\begin{document}
\hyphenpenalty=6000
\tolerance=1000

\title{Systematic study on the quark-hadron mixed phase in compact stars}
\author{Cheng-Jun~Xia$^{1,2}$}
\email{cjxia@nit.zju.edu.cn}
\author{Toshiki Maruyama$^{2}$}
\email{maruyama.toshiki@jaea.go.jp}
\author{Nobutoshi Yasutake$^{3, 2}$}
\email{nobutoshi.yasutake@it-chiba.ac.jp}
\author{Toshitaka Tatsumi$^{4}$}
\email{tatsumitoshitaka@gmail.com}
\author{Hong Shen$^{5}$}
\email{songtc@nankai.edu.cn}
\author{Hajime Togashi$^{6}$}
\email{togashi@phys.kyushu-u.ac.jp}

\affiliation{$^{1}${School of Information Science and Engineering, Zhejiang University Ningbo Institute of Technology, Ningbo 315100, China}
\\$^{2}${Advanced Science Research Center, Japan Atomic Energy Agency, Shirakata 2-4, Tokai, Ibaraki 319-1195, Japan}
\\$^{3}${Department of Physics, Chiba Institute of Technology (CIT), 2-1-1 Shibazono, Narashino, Chiba, 275-0023, Japan}
\\$^{4}${Institute of Education, Osaka Sangyo University, 3-1-1 Nakagaito, Daito, Osaka 574-8530, Japan}
\\$^{5}${School of Physics, Nankai University, Tianjin 300071, China}
\\$^{6}${Department of Physics, Kyushu University, Fukuoka, 819-0395, Japan}}

\date{\today}

\begin{abstract}
We investigate systematically the quark-hadron mixed phase in dense stellar matter, and its influence on compact star structures. The properties of quark matter and hadronic matter are fixed based on various model predictions. Beside adopting constant values, the surface tension $\Sigma$ for the quark-hadron interface is estimated with the multiple reflection expansion method and equivparticle model. To fix the structures of quark-hadron pasta phases, a continuous dimensionality of the structure is adopted as proposed by Ravenhall, Pethick, and Wilson. The corresponding properties of hybrid stars are then obtained and confronted with pulsar observations. It is found that the correlation between radius and tidal deformability in traditional neutron stars preserves in hybrid stars. For those permitted by pulsar observations, in almost all cases the quark phase persists inside the most massive compact stars. The quark-hadron interface plays an important role on hybrid star structures once quark matter emerges. The surface tension $\Sigma$ estimated with various methods increases with density, which predicts stiffer EOSs for the quark-hadron mixed phase and increases the maximum mass of hybrid stars. {With or without the emergence of quark matter, the obtained EOSs of hybrid star matter are close to each other at densities $n\lesssim 0.8$ fm${}^{-3}$}, while larger uncertainty is expected at higher densities.
\end{abstract}


\maketitle

\section{\label{sec:intro}Introduction}

Due to the asymptotic freedom of strong interaction, the deconfinement phase transition is expected as one increases the density of hadronic matter. However, it is still unclear how such a transition takes place. Traditionally, for zero temperature cases, a first-order phase transition between hadronic matter (HM) and quark matter (QM) was predicted by various quark models, which indicates a quark-hadron mixed phase (MP)~\cite{Glendenning2000, Peng2008_PRC77-065807, Li2015_PRC91-035803, Klahn2013_PRD88-085001, Bombaci2016_IJMPD-1730004}. Adopting different surface tension values for the quark-hadron interface, it was found that the MP exhibits various structures~\cite{Maruyama2007_PRD76-123015}. For vanishing surface tensions and Coulomb interactions, the MP is comprised of HM and QM that satisfy the Gibbs condition~\cite{Glendenning2000}. If a moderate surface tension is employed, with the charged particles relocate themselves via charge screening effects, geometrical structures appear~\cite{Heiselberg1993_PRL70-1355, Voskresensky2002_PLB541-93, Tatsumi2003_NPA718-359, Voskresensky2003_NPA723-291, Endo2005_NPA749-333, Maruyama2007_PRD76-123015, Yasutake2014_PRC89-065803, Xia2019_PRD99-103017, Maslov2019_PRC100-025802}. Those structures become unstable for enough large surface tensions, which leads to a bulk separation of quark and hadron phases, i.e., the Maxwell construction scenarios.

Ever since the first discovery in 1967~\cite{Hewish1968_Nature217-709}, more than 2800 pulsars have been observed~\cite{Manchester2015}. This number is increasing exponentially with the implementation of powerful telescopes~\cite{Li2015_IAUGA22-2251846, NAN2011_IJMPD20-989, Smits2009_AA493-1161, LAMP2015}. Being the natural laboratory of dense matter, the observation of pulsars has put strong constraints on the properties of strongly interacting matter at highest densities~\cite{Lattimer2012_ARNPS62-485, Ozel2016_ARAA54-401, Baiotti2019_PPNP109-103714, Weih2019_ApJ881-73}. By analyzing its orbital motion through the arrival times of the pulsations, the masses of approximately 70 pulsars in binary systems were measured~\cite{Lattimer2012_ARNPS62-485}, where the precise mass measurements of the two-solar-mass pulsars PSR J1614-2230 ($1.928 \pm 0.017\ M_\odot$)~\cite{Demorest2010_Nature467-1081, Fonseca2016_ApJ832-167} and PSR J0348+0432 ($2.01 \pm 0.04\ M_\odot$)~\cite{Antoniadis2013_Science340-1233232} have put strong constraints on the equation of state (EOS) of dense stellar matter. Pulsars that are heavier than $2\ M_{\odot}$ are expected, e.g., the presently heaviest PSR J0740+6620 ($2.14^{+0.10}_{-0.09}M_{\odot}$)~\cite{Cromartie2020_NA4-72} and possibly the more massive PSR J2215+5135 ($2.27{}_{-0.15}^{+0.17}\ M_\odot$)~\cite{Linares2018_ApJ859-54}. Nevertheless, based on the numerical simulations of binary neutron star merger event GW170817, an upper limit of the maximum mass has been suggested ($\le2.35 M_{\odot}$)~\cite{Rezzolla2018_ApJ852-L25, Ruiz2018_PRD97-021501, Shibata2019_PRD100-023015}. Both the masses and radii of pulsars may be accurately measured via pulse-profile modeling~\cite{Watts2019_SCPMA62-29503}, where recently the NICER mission has obtained the mass (1.18-$1.59\ M_\odot$) and radius (11.52-14.26 km) of PSR J0030+0451~\cite{Riley2019_ApJ887-L21, Miller2019_ApJ887-L24}. With the first observation of gravitational waves from GW170817 event, the dimensionless combined tidal deformability of pulsars are constrained within $302\leq \tilde{\Lambda} \leq 720$~\cite{LVC2017_PRL119-161101, LVC2019_PRX9-011001, Coughlin2019_MNRAS489-L91, Carney2018_PRD98-063004, De2018_PRL121-091102, Chatziioannou2018_PRD97-104036}, with the corresponding radii estimated to be $11.9^{+1.4}_{-1.4}$ km~\cite{LVC2018_PRL121-161101}. A combination of the observed masses, radii, and tidal deformabilities of pulsars gives rise to the strongest constraint for dense matter.

In our previous study~\cite{Xia2019_PRD99-103017}, we have considered the possibility of constraining the surface tension from pulsar observations, where a first-order deconfinement phase transition was assumed. By adopting the covariant density functional TW99~\cite{Typel1999_NPA656-331} for nuclear matter and perturbation model~\cite{Fraga2005_PRD71-105014} for quark matter, it was found that varying the surface tension value will have sizable effects on the radii and tidal deformabilities of 1.36-solar-mass hybrid stars.

Nevertheless, due to the important roles played by many-body interactions as well as the emergence of hadrons other than nucleons, the properties of hadronic matter at densities larger than twice the nuclear saturation density are not very well constrained, where the differences between various predictions grow dramatically~\cite{Hu2017_PRC96-034307}. Meanwhile, even though the perturbation model gives reliable predictions at ultra-high densities~\cite{Fraga2014_ApJ781-L25}, the properties of quark matter inside hybrid stars are poorly constrained. Under such circumstances, in the present work, we further extend our study by investigating systematically the hadron-quark deconfinement phase transition in dense stellar matter, where various combinations of models that describe QM and HM are adopted along with different values of surface tension.

For hadronic matter, we adopt 10 different EOSs predicted by relativistic-mean-field (RMF) model~\cite{Meng2016_RDFNS} and variational method with realistic baryon interactions~\cite{Akmal1998_PRC58-1804, Togashi2017_NPA961-78}. Among them, two EOSs include the contributions of hyperons explicitly. For the quark phase, we adopt 46 EOSs predicted by equivparticle model~\cite{Peng2000_PRC62-025801, Wen2005_PRC72-015204, Xia2014_PRD89-105027}, perturbation model~\cite{Freedman1977_PRD16-1169, Fraga2005_PRD71-105014, Kurkela2010_PRD81-105021}, and Nambu-Jona-Lasinio (NJL) model~\cite{Hatsuda1994_PR247-221, Rehberg1996_PRC53-410}.

To fix the structures of quark-hadron pasta phases, a continuous dimensionality of the structure is introduced as proposed by~\citet{Ravenhall1983_PRL50-2066}. The energy contribution due to the quark-hadron interface is treated with a surface tension $\Sigma$, for which we employ constant values as well as those estimated by the multiple reflection expansion method~\cite{Berger1987_PRC35-213, *Berger1991_PRC44-566, Madsen1993_PRL70-391, Madsen1993_PRD47-5156, Madsen1994_PRD50-3328} and equivparticle model including both linear confinement and leading-order perturbative interactions~\cite{Xia2018_PRD98-034031, Xia2019_AIPCP2127-020029}.

The EOSs of hybrid star matter are obtained, while the corresponding compact star structures are determined by solving the Tolman-Oppenheimer-Volkov (TOV) equation. For the EOSs of hybrid star matter consistent with pulsar observations, it is found that in almost all cases the quark phase takes place inside the most massive compact stars. Once quark matter emerges, we find that the quark-hadron interface plays an important role on hybrid star structures.

The paper is organized as follows. We present our theoretical framework in Sec.~\ref{sec:theHM}, Sec.~\ref{sec:theQM}, and Sec.~\ref{sec:MP}. Two formalisms are adopted for the HM, i.e., the RMF model in Sec.~\ref{sec:the_RMF} and the variational method in Sec.~\ref{sec:the_VM}. The equivparticle model, perturbation model, and NJL model for QM are introduced in Sec.~\ref{sec:theQM}. The formalism in obtaining the structures of quark-hadron mixed phase is introduced in Sec.~\ref{sec:pasta}, while the surface tension of quark-hadron interface is obtained in Sec.~\ref{sec:the_surf}. The numerical results are presented and discussed in Sec.~\ref{sec:star}. Our conclusion is given in Sec.~\ref{sec:con}.

\section{\label{sec:theHM}Effective models for hadronic matter}

\subsection{\label{sec:the_RMF} RMF model}
The Lagrangian density for infinite nuclear matter obtained with RMF model~\cite{Meng2016_RDFNS} reads
\begin{eqnarray}
\mathcal{L}_\mathrm{NM} &=& \sum_{i=n,p} \bar{\Psi}_i
       \left[  i \gamma^\mu \partial_\mu - m^*   -\gamma^0 \left(g_{\omega} \omega
               + g_{\rho} \tau_{i,3} \rho_3\right)\right] \Psi_i
\nonumber \\
&&\mbox{}
     - \frac{1}{2}m_\sigma^2 \sigma^2
     + \frac{1}{2}m_\omega^2 \omega^2
     + \frac{1}{2}m_\rho^2 \rho_3^2+U(\sigma, \omega). \label{eq:Ll}
\end{eqnarray}
Here the Dirac spinor $\Psi_i$ represents nucleons with the effective mass $m^* = m+g_\sigma \sigma$ and isospin $\boldsymbol{\tau}_i$. Three types of mesons are included to describe the interactions between nucleons, i.e., $\sigma$-, $\omega$-, and $\rho$-mesons with their masses being $m_\sigma$, $m_\omega$ and $m_\rho$, respectively. The baryon number density is given by $n= n_n + n_p = \sum_{i=n, p} \langle\bar{\Psi}_i \gamma^0 \Psi_i\rangle$. In this work, we adopt two different schemes for the density dependence of effective interaction strengths, i.e., the nonlinear self-couplings of $\sigma$ and $\omega$ mesons in $U(\sigma, \omega)$ and the Typel-Wolter ansatz with density dependent coupling constants~\cite{Typel1999_NPA656-331}.

The nonlinear self-couplings for $\sigma$ and $\omega$ mesons are
\begin{equation}
  U(\sigma, \omega) =  - \frac{1}{3} g_2 \sigma^3 - \frac{1}{4} g_3 \sigma^4 +\frac{1}{4} c_3 \omega^4,
\end{equation}
where we have adopted the effective interaction TM1~\cite{Sugahara1994_NPA579-557}, i.e., Shen EOS2~\cite{Shen2011_ApJ197-20}. Meanwhile, it was shown that the slope of symmetry energy $L = 110.8$ MeV predicted by TM1 was too large according to various constrains from nuclear physics and pulsar observations, which can be reduced to $L = 40$ MeV by adding the cross coupling term
\begin{equation}
\mathcal{L}_{\omega\rho}=\Lambda_\mathrm{v}g_\omega^2 g_\rho^2 \omega^2\rho^2.
\end{equation}
This gives Shen EOS4 by adopting the effective interaction TM1e~\cite{Shen2020_ApJ891-148}. To further include the contribution of $\Lambda$ hyperons, in Eq.~(\ref{eq:Ll}) we add the following Lagrangian density~\cite{Wang2013_CTP60-479, Lu2011_PRC84-014328, Hagino2014_arXiv1410.7531, Sun2018_CPC42-25101},
\begin{eqnarray}
\mathcal{L}_{Y}&=&\bar{\psi}_{\Lambda}
       \left[i \gamma^\mu \partial_\mu- m_\Lambda^* - \gamma^0 \alpha_{\omega\Lambda}g_{\omega} \omega\right]\psi_{\Lambda},
              \label{eq:Lagrange_LN}
\end{eqnarray}
where $m_\Lambda^* = m_{\Lambda} + \alpha_{\sigma\Lambda} g_{\sigma}\sigma$ is the effective mass of the $\Lambda$ hyperon. The ratio of coupling constants $\alpha_{\omega\Lambda}\equiv g_{\omega\Lambda}/g_{\omega} = 2/3$ is predicted by the naive quark model~\cite{Dover1984_PPNP12-171}, then $\alpha_{\sigma\Lambda}\equiv g_{\sigma\Lambda}/g_{\sigma} = 0.621$ is obtained by reproducing the binding energies of $\Lambda$-hyperon in $\Lambda$-hypernuclei, i.e., Shen EOS3~\cite{Shen2011_ApJ197-20}. However, the obtained hyperonic EOS is too soft to support massive neutron stars, i.e., the hyperon puzzle. To resolve this, larger values of $g_{\omega\Lambda}$ were adopted to provide more repulsive interaction from $\omega$ meson, where in this work (denoted as TM1$\Lambda$) we take $\alpha_{\omega\Lambda}=1$ and $\alpha_{\sigma\Lambda} = 0.887$~\cite{Sun2018_CPC42-25101}. {A through investigation on such choices can be found in Ref.~\cite{Fortin2017_PRC95-065803}.}

Despite the great successes in describing finite nuclei with nonlinear self-couplings of mesons, a direct extension of the density functional to higher densities may cause problems of stability. Alternatively, we can adopt couplings that depend explicitly on densities, which can be derived from self-energies of Dirac-Brueckner calculations of nuclear matter~\cite{Typel1999_NPA656-331, Roca-Maza2011_PRC84-054309}. We thus adopt the effective nucleon-nucleon interactions PKDD~\cite{Long2004_PRC69-034319}, TW99~\cite{Typel1999_NPA656-331}, DDME2~\cite{Lalazissis2005_PRC71-024312}, and DD2~\cite{Typel2010_PRC81-015803}, where $U(\sigma, \omega) = 0$ and the density dependence of coupling constants $g_{\sigma, \omega, \rho}$~\cite{Typel1999_NPA656-331} are obtained with
\begin{eqnarray}
g_{\sigma, \omega}(n) &=& g_{\sigma, \omega}(n_0) a_{\sigma, \omega} \frac{1+b_{\sigma, \omega}(n/n_0+d_{\sigma, \omega})^2}
                          {1+c_{\sigma, \omega}(n/n_0+e_{\sigma, \omega})^2}, \label{eq:ddcp_TW}\\
g_{\rho}(n) &=& g_{\rho}(n_0) \exp{\left[-a_\rho(n/n_0-1)\right]}. \label{eq:ddcp_rho}
\end{eqnarray}
Here $n_0$ represents the nuclear saturation density.

Carrying out standard mean-field and no-sea approximations, one obtains the energy density $E$, chemical potentials $\mu_\mathrm{B}$, and pressure $P$ at given baryon density $n$. Then the EOSs for nuclear matter and hypronic matter can be obtained.

\subsection{\label{sec:the_VM} Variational methods}
The variational method for uniform nuclear matter was developed in Refs.~\cite{Kanzawa2007_NPA791-232, Kanzawa2009_PTP122-673, Togashi2013_NPA902-53}, where the nuclear Hamiltonian composed of a two-body potential $V_{ij}$ and three-body potentials $V_{ijk}$ are given by
\begin{equation}
H = -\sum_{i=1}^N \frac{\hbar^2}{2m_n}\nabla_i^2 + \sum_{i<j}^N V_{ij} + \sum_{i<j<k}^N V_{ijk}. \label{eq:VMHamiltonian}
\end{equation}
Adopting the Argonne v18 (AV18) two-body nuclear potential~\cite{Wiringa1995_PRC51-38} and the Urbana IX (UIX) three-body nuclear force~\cite{Carlson1983_NPA401-59, Pudliner1995_PRL74-4396}, the free energy per nucleon of uniform nuclear matter is predicted by the cluster variational method using the Jastrow wave function~\cite{Togashi2017_NPA961-78}. Then the equation of states for nuclear matter (denoted as VM) can be obtained, which was discussed in detail in Ref.~\cite{Togashi2017_NPA961-78}. For hyperonic EOS (VM$\Lambda$), we adopt the results presented in Ref.~\cite{Togashi2016_PRC93-035808} with three body forces of hyperons.

A more sophisticated variational method with the Fermi Hypernetted Chain calculations was performed for symmetric nuclear matter (SNM) and pure neutron matter (PNM) by Akmal, Pandharipande, and Ravenhall (APR)~\cite{Akmal1998_PRC58-1804}, where the aforementioned realistic nuclear Hamiltonian and Jastrow wave function were adopted. The energy density of nuclear matter obtained in Ref.~\cite{Akmal1998_PRC58-1804} are fixed by the fitted formula
\begin{eqnarray}
  E_\mathrm{HM} &=& \left[\frac{\hbar^2}{2m} + \left(p_3 + \frac{1+\delta}{2} p_5\right)n \mathrm{e}^{-p_4 n}   \right]\frac{\nu_n^5}{5\pi^2}   \nonumber \\
    &&{}  + \left[\frac{\hbar^2}{2m}+ \left(p_3 + \frac{1-\delta}{2} p_5\right)n \mathrm{e}^{-p_4 n}   \right]\frac{\nu_p^5}{5\pi^2} \nonumber \\
    &&{}  + g(n, \delta=0)\left(1-\delta\right)^2 + g(n, \delta=1)\delta^2.
\end{eqnarray}
Here $\delta = (n_n - n_p)/n$ is the isospin asymmetry and $\nu_{p,n}=(3\pi^2 n_{p,n})^{1/3}$ the Fermi momentum of nucleons. A more detailed description on the parameters $p_i$ and functional form $g(n, \delta)$ can be found in the original publication~\cite{Akmal1998_PRC58-1804}.  In this work, we adopt the most comprehensive case employing the AV18 two-body nuclear potential and UIX three-body interaction~\cite{Carlson1983_NPA401-59, Pudliner1995_PRL74-4396} with relativistic corrections.

\subsection{\label{sec:the_NM} The EOSs of nuclear/hyperonic matter}
Finally, for the hadronic phase, we adopt in total 10 different EOSs, i.e., 8 nuclear EOSs (TM1e, TM1, PKDD, TW99, DDME2, DD2, VM, APR) and 2 hyperonic EOSs (TM1$\Lambda$ and VM$\Lambda$). These EOSs are predicted by both RMF model with various effective interactions and variational methods started from realistic baryon interactions. The corresponding saturation properties are indicated in Table~\ref{table:NM} and compared with the constraints from terrestrial experiments and nuclear theories~\cite{Dutra2014_PRC90-055203}, which give the binding energy $\varepsilon\approx 16$ MeV, the incompressibility $K = 240 \pm 20$ MeV~\cite{Shlomo2006_EPJA30-23}, the symmetry energy $S = 31.7 \pm 3.2$ MeV and its slope $L = 58.7 \pm 28.1$ MeV~\cite{Li2013_PLB727-276, Oertel2017_RMP89-015007} around $n_0\approx 0.15\text{-}0.16\ \mathrm{fm}^{-3}$ and $\delta=0$. The uncertainties may be further reduced if the constraints from the GW170817 binary neutron star merger event~\cite{LVC2017_PRL119-161101, LVC2018_PRL121-161101} are included~\cite{Tsang2019_PLB795-533}, e.g., a recent estimation suggests $K = 250.23 \pm 20.16$ MeV, $S = 31.35 \pm 2.08$ MeV and $L = 59.57  \pm 10.06$ MeV~\cite{Zhang2020_PRC101-034303}. In general, TM1 and PKDD slightly overestimate $K$, $S$, and $L$, while TM1e predicts reasonable symmetry energy properties. The VM EOS has the smallest $S$ and $L$ but still lie within the permitted ranges.

\begin{table}
\caption{\label{table:NM} The saturation properties of nuclear matter and the corresponding maximum masses $M_\mathrm{max}$ and radii $R_{1.4}$ of 1.4 solar-mass neutron stars predicted by various methods. For TM1$\Lambda$~\cite{Sun2018_CPC42-25101} and VM$\Lambda$~\cite{Togashi2016_PRC93-035808}, the hyperons have little impact on $R_{1.4}$, while the maximum masses are reduced to 2.06 and 2.16 $M_\odot$, respectively. }
\begin{tabular}{c|ccccc|cc} \hline \hline
                                          & $n_0$        &   $\varepsilon$    &   $K$  &  $S$   & $L$   & $M_\mathrm{max}$  & $R_{1.4}$   \\
                                          & fm${}^{-3}$  &   MeV    &   MeV  &   MeV  &  MeV  & $M_\odot$         & km    \\   \hline
TM1e~\cite{Shen2020_ApJ891-148}           &  0.145       &  16.26   & 281.16 &  31.38 & 40    &  2.13             & 13.1 \\
TM1~\cite{Sugahara1994_NPA579-557}        &  0.145       &  16.26   & 281.16 &  36.89 & 110.79&  2.18             & 14.3 \\
PKDD~\cite{Long2004_PRC69-034319}         &  0.150       &  16.27   & 262.19 &  36.79 & 90.21 &  2.33             & 13.6 \\
TW99~\cite{Typel1999_NPA656-331}          &  0.153       &  16.25   & 240.27 &  32.77 & 55.31 &  2.08             & 12.3 \\
DDME2~\cite{Lalazissis2005_PRC71-024312}  &  0.152       &  16.14   & 250.92 &  32.30 & 51.25 &  2.49             & 13.2 \\
DD2~\cite{Typel2010_PRC81-015803}         &  0.149       &  16.02   & 242.72 &  31.67 & 55.04 &  2.43             & 12.8  \\
VM~\cite{Togashi2017_NPA961-78}           &  0.160       &  16.09   & 245    &  30.0  & 37    &  2.22             & 11.6 \\
APR~\cite{Akmal1998_PRC58-1804}           &  0.160       &  16.00   & 269.28 &  33.94 & 57.9  &  2.19             & 11.4 \\
\hline
\end{tabular}
\end{table}

\begin{figure}
\includegraphics[width=\linewidth]{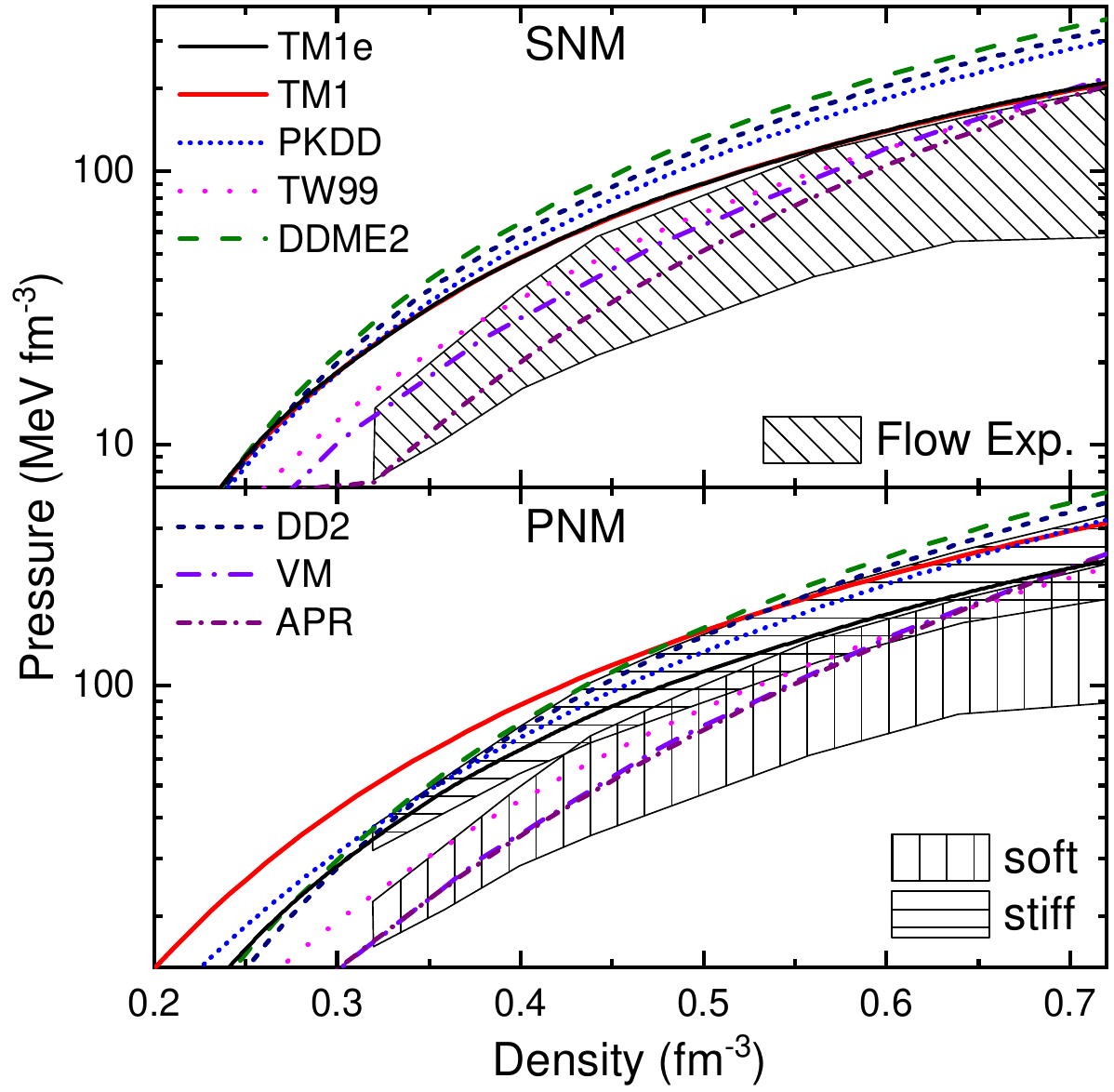}
\caption{\label{Fig:Flow} The pressures of symmetric nuclear matter (SNM) and pure neutron matter (PNM) predicted by various nuclear theories, which are compared with the experimental constraints from the flow data~\cite{Danielewicz2002_Science298-1592}. }
\end{figure}

At larger densities, in Fig.~\ref{Fig:Flow} we present the pressures of SNM and PNM as functions of baryon number density, which are compared with the constraints from the flow data of heavy ion collisions~\cite{Danielewicz2002_Science298-1592}. It is found that the EOSs of nuclear matter predicted by TM1e, TM1, PKDD, DDME2 and DD2 are slightly stiffer than those constrained from the flow data of heavy ion collisions~\cite{Danielewicz2002_Science298-1592}. Nevertheless, the emergence of the quark phase may ease the tension and reduce the stiffness of EOSs effectively.

\begin{figure}
\includegraphics[width=\linewidth]{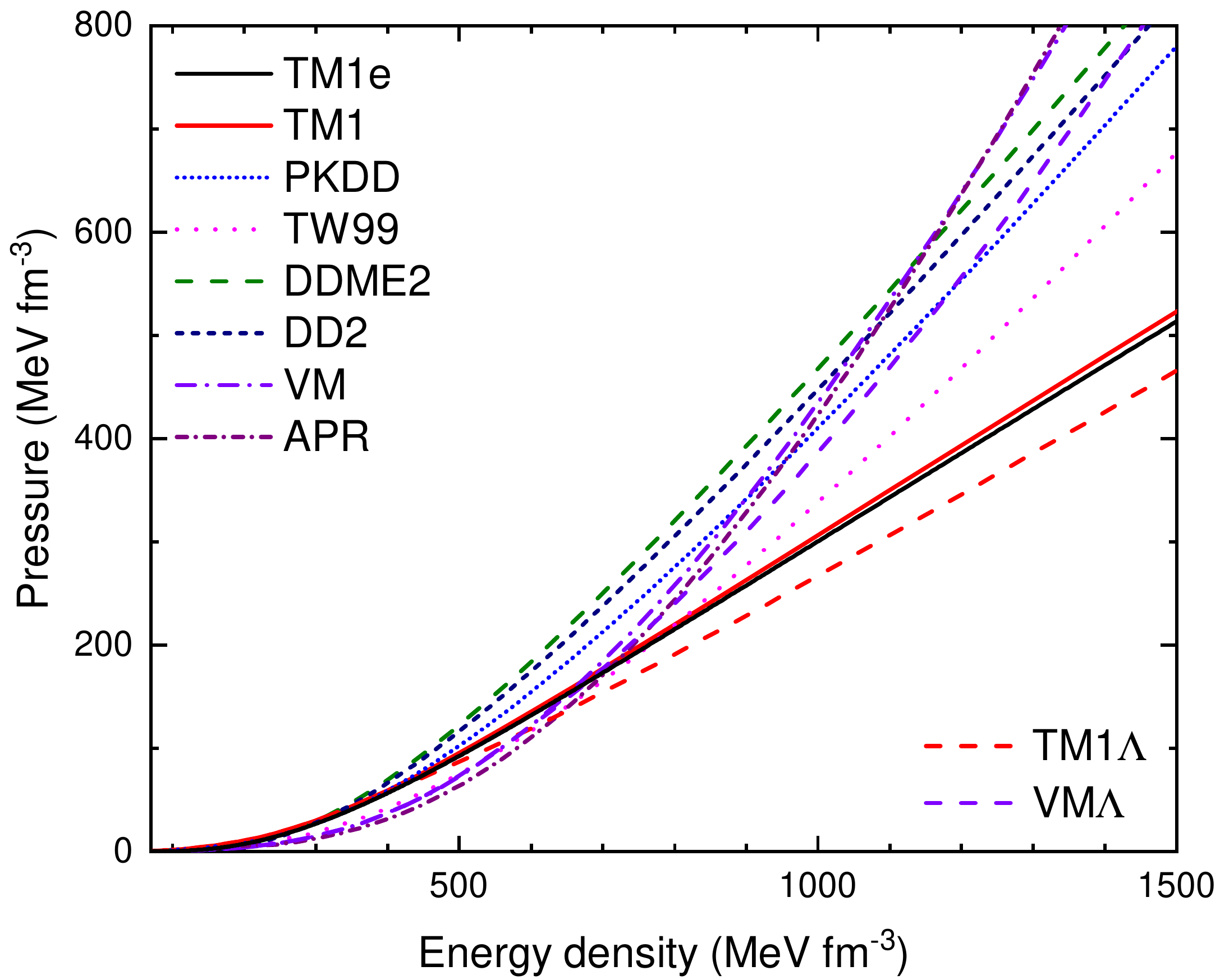}
\caption{\label{Fig:HMEoS} The pressure of neutron star matter as functions of energy density. }
\end{figure}

The EOS of neutron star matter can be obtained by further including the contributions of electrons and muons, where their energy densities take the form of free Fermi gas with
\begin{equation}
E_{i}^0 = \frac{g_i m_i^4}{16\pi^{2}} \left[x_i(2x_i^2+1)\sqrt{x_i^2+1}-\mathrm{arcsh}(x_i) \right]. \label{eq:E0}
\end{equation}
Here $g_{e,\mu}=2$ is the degeneracy factor and $x_{e,\mu}\equiv \nu_{e,\mu}/m_{e,\mu}$ with $\nu_{e,\mu}$ being the Fermi momentum of leptons, which predicts their number densities $n_{e,\mu} = \nu_{e,\mu}^3/3\pi^2$. The total energy density of neutron star matter is obtained with $E=E_\mathrm{HM}+E_{e}+E_{\mu}$. Then the pressure is determined by $P = \sum_i \mu_i n_i  - E$ with the chemical potential $\mu_i = \frac{\partial E}{\partial n_i}$. In Fig.~\ref{Fig:HMEoS} we present the EOSs of neutron star matter, which are obtained by simultaneously fulfilling the $\beta$-stability condition and local charge neutrality condition.

\begin{figure}
\includegraphics[width=\linewidth]{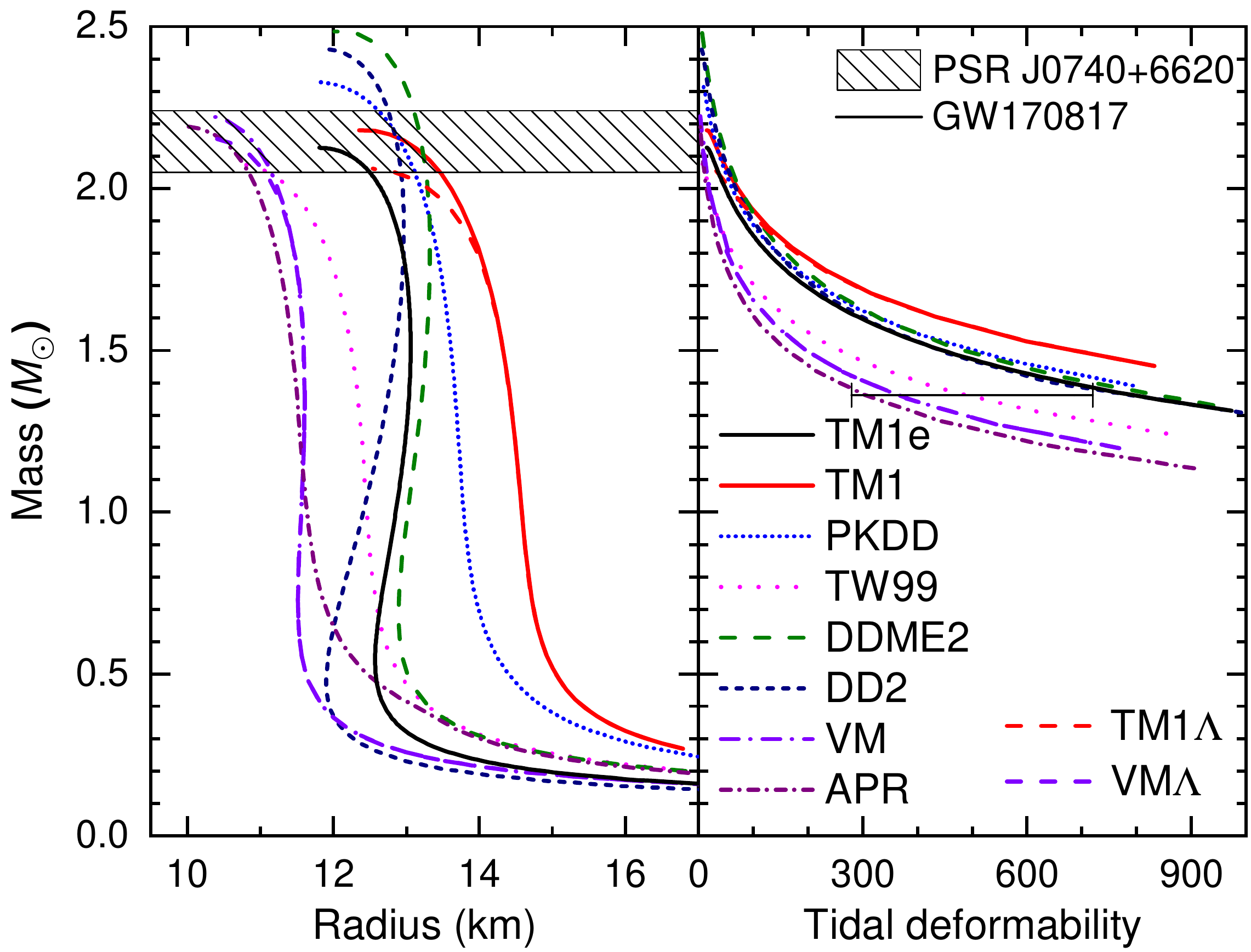}
\caption{\label{Fig:MLH} The mass, radius, and tidal deformability of neutron stars obtained with the EOSs presented in Fig.~\ref{Fig:HMEoS}. The maximum masses $M_\mathrm{max}$ and radii $R_{1.4}$ of $1.4 M_\odot$ neutron stars are indicated in Table~\ref{table:NM}.}
\end{figure}

Based on the EOSs indicated in Fig.~\ref{Fig:HMEoS}, the corresponding structures of compact stars are obtained by solving the TOV equation
\begin{eqnarray}
&&\frac{\mbox{d}P}{\mbox{d}r} = -\frac{G M E}{r^2}   \frac{(1+P/E)(1+4\pi r^3 P/M)} {1-2G M/r},  \label{eq:TOV}\\
&&\frac{\mbox{d}M}{\mbox{d}r} = 4\pi E r^2, \label{eq:m_star}
\end{eqnarray}
while the tidal deformability is estimated with
\begin{equation}
\Lambda = \frac{2 k_2}{3}\left( \frac{R}{G M} \right)^5. \label{eq:td}
\end{equation}
Here the gravity constant is taken as $G=6.707\times 10^{-45}\ \mathrm{MeV}^{-2}$, while $k_2$ is the second Love number and is obtained from the response of the induced quadrupole moment ${\cal Q}_{ij}$ in a static external quadrupolar tidal field ${\cal E}_{ij}$ with ${\cal Q}_{ij} = -k_2 \frac{2R^5}{3G}{\cal E}_{ij}$~\cite{Damour2009_PRD80-084035, Hinderer2010_PRD81-123016, Postnikov2010_PRD82-024016}. Note that at $n<0.08\ \mathrm{fm}^{-3}$ we have adopted the EOSs presented in Refs.~\cite{Feynman1949_PR75-1561, Baym1971_ApJ170-299, Negele1973_NPA207-298}, which account for the crusts of neutron stars. For the cases of TM1, TM1$\Lambda$, TM1e, VM, and VM$\Lambda$, the crust EOSs were previously obtained, i.e., Shen EOSs~\cite{Shen2011_ApJ197-20, Shen2020_ApJ891-148} and VM EOSs~\cite{Togashi2016_PRC93-035808, Togashi2017_NPA961-78}. However, instead of using those EOSs, we still adopt the crust EOSs presented in Refs.~\cite{Feynman1949_PR75-1561, Baym1971_ApJ170-299, Negele1973_NPA207-298} since the variations on neutron star structures are relatively small. The obtained mass, radius, and tidal deformability are presented in Fig.~\ref{Fig:MLH} and compared with astrophysical observations, where the maximum masses and radii of $1.4 M_\odot$ neutron stars are indicated in Table~\ref{table:NM}. All the maximum masses of compact stars predicted by various EOSs in Fig.~\ref{Fig:HMEoS} are consistent with the observational mass ($2.14^{+0.10}_{-0.09}M_{\odot}$) of PSR J0740+6620~\cite{Cromartie2020_NA4-72}. Nevertheless, we should mention that the velocity of sound will exceed $c$ for APR, VM, and VM$\Lambda$ at $n\geq 0.87$, 0.90, and 1.08 fm$^{-3}$, which are reached in center regions of the massive compact stars indicated in Fig.~\ref{Fig:MLH}. The tidal deformabilities predicted by TM1e, PKDD, DDME2, and DD2 slightly exceed the constraint $302\leq \tilde{\Lambda} \leq 720$ from the GW170817 binary neutron star merger event~\cite{LVC2017_PRL119-161101, LVC2019_PRX9-011001, Coughlin2019_MNRAS489-L91, Carney2018_PRD98-063004, De2018_PRL121-091102, Chatziioannou2018_PRD97-104036}, which coincide with the experimental constraints on SNM from the flow data~\cite{Danielewicz2002_Science298-1592} in Fig.~\ref{Fig:Flow}. Note that the recent radius measurements of PSR J0030+0451 with the equatorial radius $R_\mathrm{eq}=11.52$-14.26 km and mass $M = 1.18$-$1.59\ M_\odot$ obtained via pulse-profile modeling in the NICER mission~\cite{Riley2019_ApJ887-L21, Miller2019_ApJ887-L24} do not constrain the EOSs adopted here.

\section{\label{sec:theQM}Effective models for quark matter}
\subsection{\label{sec:the_equiv}Equivparticle model}
In the equivparticle model, the quarks are treated as quasi-free particles with density dependent equivalent masses. Taking into account both the linear confinement and leading order perturbative interactions, the quark mass scaling is given by~\cite{Xia2014_PRD89-105027}
\begin{equation}
  m_i(n)=m_{i0}+\frac{D}{n^{1/3}}+Cn^{1/3}.  \label{eq:mnbC}
\end{equation}
Here $m_{i0}$ is the current mass of quark flavor $i$ ($i=u, d, s$)~\cite{Olive2014_CPC38-090001} and $n\equiv \sum_{i=u, d, s} n_i/3$ the baryon number density. The confinement parameter $D$ is connected to the string tension $\sigma_0$, the chiral restoration density $\rho^*$, and the sum of the vacuum chiral condensates $\sum_q\langle\bar{q}q\rangle_0$. Meanwhile, the perturbative strength parameter $C$ is linked to the strong coupling constant $\alpha_\mathrm{s}$. Due to the uncertainties in relevant quantities, we do not know the exact values of $D$ and $C$. Nevertheless, it has been estimated that $\sqrt{D}$ approximately lies in the range of (147, 270) MeV~\cite{Wen2005_PRC72-015204} and $C\lesssim 1.2$~\cite{Xia2014_PRD89-105027}. In this work, we adopt the parameter sets ($C$, $\sqrt{D}$ in MeV): ($-0.2$, 180), (0, 180), (0.7, 140), (0.7, 180), (1, 140), and (1, 180).

At zero temperature, the energy density $E_\mathrm{QM} = \sum_{i=u, d, s} E_{i}^0(\nu_i, m_i)$ and particle number density $n_i =  {\nu_i^3}/{\pi^2}$ are identical to the cases of free Fermi gas with $E_{i}^0$ given by Eq.~(\ref{eq:E0}) and $g_{u, d, s} = 6$. Note that in Eq.~(\ref{eq:E0}) we have adopted the mass scaling of Eq.~(\ref{eq:mnbC}) for quarks, i.e., $m_i\equiv m_i(n)$. The pressure is determined by $P_\mathrm{QM} = \sum_{i=u, d, s} \mu_i n_i  - E_\mathrm{QM}$ with the chemical potential $\mu_i = \frac{\partial E_\mathrm{QM}}{\partial n_i}$.

\subsection{\label{sec:the_pQCD} Perturbation model}
By expanding the thermodynamic potential density of quark matter to the order of $\alpha_\mathrm{s}$~\cite{Fraga2005_PRD71-105014}, one obtains
\begin{equation}
\Omega^\mathrm{pt} = \sum_i^{N_f} \left( \omega^0_i + \omega^1_i \alpha_\mathrm{s} \right), \label{eq:omega}
\end{equation}
with
\begin{eqnarray}
\omega^0_i &=& - \frac{g_i m_i^4}{24 \pi^2}
                 \left[u_i v_i \left( u_i^2 - \frac{5}{2}  \right)
                   + \frac{3}{2} \ln(u_i+v_i)
                 \right],
\label{eq:omega0}\\
\omega^1_i &=& \frac{g_i m_i^4}{12\pi^3}
               \left\{ \left[ 6 \ln\left(\frac{\bar{\Lambda}}{m_i}\right) + 4 \right]\left[u_i v_i - \ln(u_i+v_i)\right] \right.
\nonumber\\
          &&\left.
               + \mbox{} 3\left[u_i v_i - \ln(u_i+v_i)\right]^2 - 2 v_i^4 \right\},
\label{eq:omega1}
\end{eqnarray}
where $u_i \equiv \mu_i/m_i$ and $v_i \equiv \sqrt{u_i^2-1}$ with $\mu_i$ and $m_i$ being the chemical potential and mass of quark flavor $i$. The running coupling constant and quark
masses are obtained by solving the $\beta$-function and $\gamma$-function~\cite{Vermaseren1997_PLB405-327}, which gives~\cite{Fraga2005_PRD71-105014}
\begin{eqnarray}
\alpha_\mathrm{s}(\bar{\Lambda})
  &=& \frac{1}{\beta_0 L}   \left(1- \frac{\beta_1\ln{L}}{\beta_0^2 L}\right),
\label{eq:alpha} \\
m_i(\bar{\Lambda})
  &=& \hat{m}_i \alpha_\mathrm{s}^{\frac{\gamma_0}{\beta_0}}
      \left[ 1 + \left(\frac{\gamma_1}{\beta_0}-\frac{\beta_1\gamma_0}{\beta_0^2}\right) \alpha_\mathrm{s} \right].
\label{eq:mi}
\end{eqnarray}
Here $L=2 \ln\left( \frac{\bar{\Lambda}}{\Lambda_{\overline{\mathrm{MS}}}}\right)$ with $\Lambda_{\overline{\mathrm{MS}}} = 376.9$ MeV being the $\overline{\mathrm{MS}}$ renormalization point, while the invariant quark masses are fixed as $\hat{m}_u= 3.8$ MeV, $\hat{m}_d = 8$ MeV, and $\hat{m}_s = 158$ MeV~\cite{Olive2014_CPC38-090001}. The parameters are given by $\beta_0=9/4\pi$, $\beta_1=4/\pi^2$, $\gamma_0=1/\pi$ and $\gamma_1=91/24 \pi^2$. In this work, we take the renormalization scale $\bar{\Lambda} = C_1(\mu_u+\mu_d+\mu_s)/3$ with $C_1=1$--4~\cite{Fraga2014_ApJ781-L25}, while a parameterized bag constant is also adopted~\cite{Burgio2002_PLB526-19, Maieron2004_PRD70-043010, Xia2019_PRD99-103017}, i.e.,
\begin{equation}
B = B_\mathrm{QCD} + (B_0 - B_\mathrm{QCD})
    \exp{\left[-\left( \frac{\sum_i\mu_i-930}{\Delta\mu}\right)^4\right]}
\label{eq:BL}
\end{equation}
with $B_\mathrm{QCD} = 400\ \mathrm{MeV\ fm}^{-3}$ and $B_0 = 50\ \mathrm{MeV\ fm}^{-3}$. Finally, the thermodynamic potential density for quark matter is given by $\Omega_\mathrm{QM} =  \Omega^\mathrm{pt} + B$. The particle number density, energy density, and pressure are then obtained with $n_i = - \frac{\partial \Omega_\mathrm{QM}}{\partial \mu_i}$, $E_\mathrm{QM} =  \Omega_\mathrm{QM} + \sum_i \mu_i n_i$, and $P_\mathrm{QM}=-\Omega_\mathrm{QM}$. In this work, we take the parameters $C_1 = 2$, 2.5, 3, 3.5 and $\Delta\mu = 770$, 800, 830, 860, 890, 920, 950, 980 MeV.

\subsection{\label{sec:NJL} NJL model with vector interactions}
In the mean-field approximation, the Lagrangian density of a SU(3) NJL model is given by
\begin{eqnarray}
  L_\mathrm{NJL}&=& \sum_{i = u,d,s} \bar{\psi}_i\left[ i\gamma^\mu \partial_\mu - M_i -4 G_V \gamma^0 n_i  \right]\psi_i \label{eq:Lgrg_NJL}\\
   &&{}  + 2 \sum_{i = u,d,s} \left(G_V n_i^2 - G_S \sigma_i^2\right) + 4 K\sigma_u \sigma_d \sigma_s, \nonumber
\end{eqnarray}
where the constituent quark mass reads
\begin{equation}
  M_i = m_{i0} - 4 G_S\sigma_i + 2K \sigma_j\sigma_k.
\end{equation}
Note that in Eq.~(\ref{eq:Lgrg_NJL}) a term in the vector-isoscalar channel is included, which provides repulsive interactions with $G_V>0$~\cite{Buball2005_PR407-205}.

At $T = 0$, the thermodynamic potential density of quark matter predicted by the NJL model is determined by
\begin{eqnarray}
\Omega_\mathrm{QM} &=& \sum_{i = u,d,s}[ \omega^0_i(\mu_i^*, M_i) - E_{i}^0(\Lambda, M_i)  + 2 G_S \sigma_i^2 \nonumber\\
                    &&{} - 2 G_V n_i^2 ]  - 4 K\sigma_u \sigma_d \sigma_s - E_0 \label{eq:Omega_NJL}
\end{eqnarray}
with $E_{i}^0$ ($x_i=\Lambda/M_i$) given by Eq.~(\ref{eq:E0}) and $\omega^0_i$ ($u_i=\mu_i^* /M_i$) by Eq.~(\ref{eq:omega0}). Here a constant $E_0$ is introduced to ensure $\Omega_\mathrm{QM} = 0$ in the vacuum. $\Lambda$ is the three dimensional momentum cutoff to regularize the vacuum part, and $\mu_i^*$ the effective chemical potential which is connected with the true chemical potential via
\begin{equation}
  \mu_i^* =  \mu_i - 4G_Vn_i.
\end{equation}
Based on the thermodynamic potential density in Eq.~(\ref{eq:Omega_NJL}), the chiral condensate is given by $\sigma_i=\frac{\partial\Omega_\mathrm{QM}}{\partial M_i}$ and quark number density $n_i =  {\nu_i^3}/{\pi^2}$ with $\nu_i^2=(\mu_i^*)^2-M_i^2$. At fixed $\mu_i^*$, the equations for the chiral condensate $\sigma_i$, quark number density $n_i$, and constituent quark mass $M_i$ are solved in an iterative manner. The energy density and pressure are then obtained with $E_\mathrm{QM} =  \Omega_\mathrm{QM} + \sum_i \mu_i n_i$ and $P_\mathrm{QM}=-\Omega_\mathrm{QM}$. In this work, two different sets of parameters are adopted, i.e., the sets HK ($\Lambda = 631.4$ MeV, $m_{u0}= m_{d0} = 5.5$ MeV,  $m_{s0}=  135.7$ MeV, $G_S = 1.835/\Lambda^2$, $K = 9.29/\Lambda^5$)~\cite{Hatsuda1994_PR247-221} and RKH ($\Lambda = 602.3$ MeV, $m_{u0}= m_{d0} = 5.5$ MeV,  $m_{s0}=  140.7$ MeV, $G_S = 1.835/\Lambda^2$, $K = 12.36/\Lambda^5$)~\cite{Rehberg1996_PRC53-410}. For the vector coupling $G_V$, the Fierz-transition predicts $G_V=0.5G_S$, while in this work we take it as a free parameter with $G_V=0$, 0.5$G_S$, $G_S$, and 1.5$G_S$.

\subsection{\label{sec:the_QM} General discussion on the quark EOSs}
In contrast to nuclear matter cases, we have little constraints on the properties of quark matter at intermediate densities. At ultra-high densities ($n \gtrsim 40n_0$), however, Quantum Chromodynamics (QCD) can be solved with perturbative approaches~\cite{Fraga2014_ApJ781-L25}. The corresponding EOS at highest densities is then expected to be reproduced by the perturbation model (pQCD) explained in Sec.~\ref{sec:the_pQCD}. The NJL model, on the other hand, is a low-energy model for QCD, where the gluons are integrated out while retaining only local quark interactions. The corresponding coupling constants of NJL model are then fixed by reproducing the masses of $\pi$, $K$, $\eta'$ and the $\pi$ decay constant~\cite{Hatsuda1994_PR247-221, Rehberg1996_PRC53-410}. The equivparticle model carries similar traits of quasiparticle model~\cite{Wen2012_PRD86-034006, Plumari2011_PRD84-094004}, where the results of pQCD at highest densities can be reproduced with the parameter $C$ in Eq.~(\ref{eq:mnbC}) depending explicitly on $\alpha_\mathrm{s}$~\cite{Xia2014_PRD89-105027}. Meanwhile, the linear confinement of quarks are well treated with an inversely cubic mass scaling in the equivparticle model~\cite{Peng1999_PRC61-015201}. Note that in this work we have neglected the effects of color superconductivity~\cite{Alford2008_RMP80-1455}, which shall be considered in our future studies.

\begin{figure}
\includegraphics[width=\linewidth]{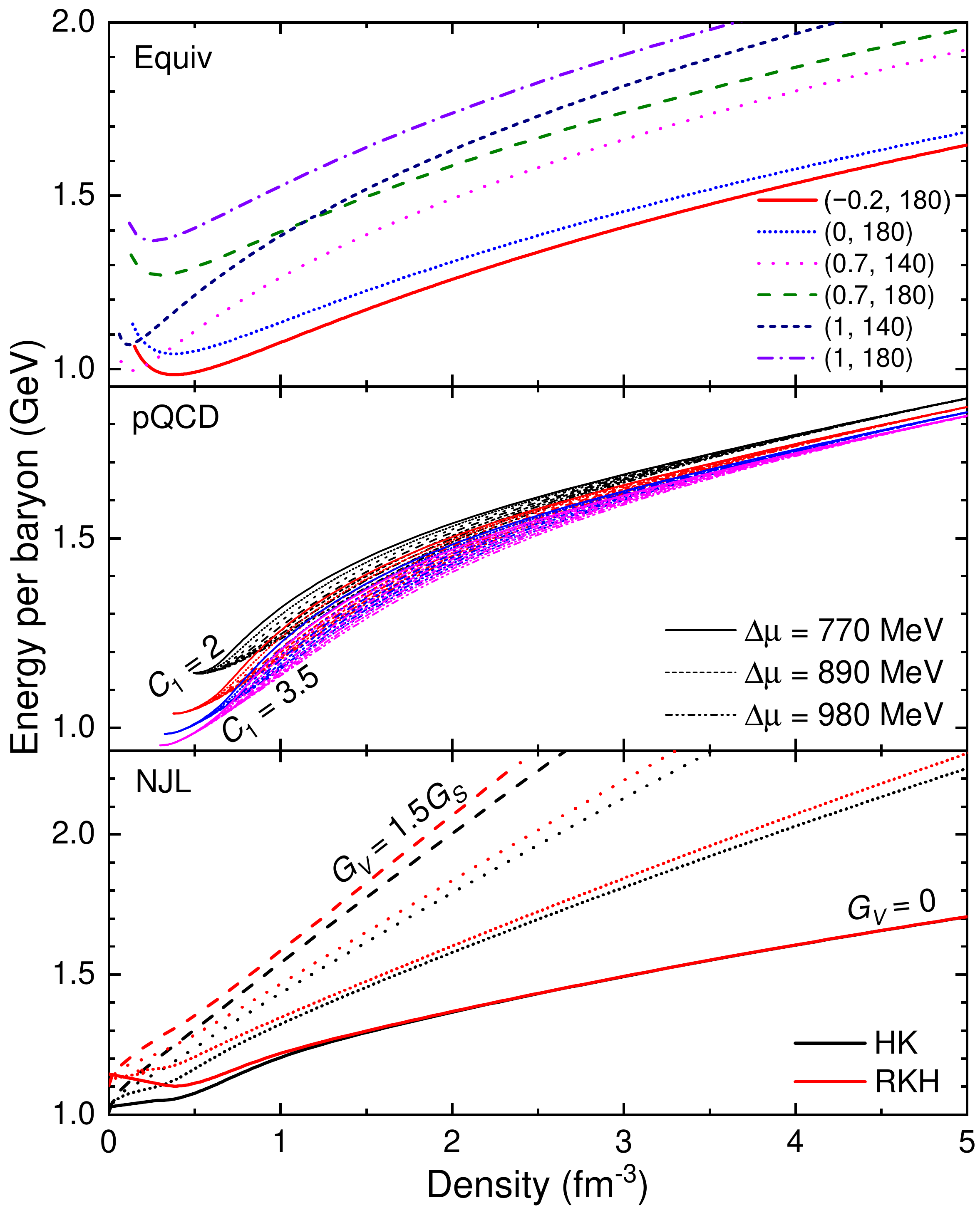}
\caption{\label{Fig:QMEoS} The energy per baryon of quark matter as functions of baryon number density $n$ predicted by equivparticle, perturbation, and NJL models. For the equivparticle model (upper panel), the parameter sets ($C$, $\sqrt{D}$) are indicated explicitly. }
\end{figure}

With the energy contributions of leptons determined by Eq.~(\ref{eq:E0}), the EOSs of quark matter in compact stars can be obtained by simultaneously fulfilling the $\beta$-stability condition $\mu_u+\mu_e=\mu_d=\mu_s$, $\mu_e=\mu_\mu$ and local charge neutrality condition $\sum_i q_i n_i=0$ with $q_i$ ($q_u=2/3$, $q_d=q_s=-1/3$ and $q_e=q_\mu=-1$) being the charge of particle type $i$. The corresponding energy per baryon of quark matter predicted by various quark models are then presented in Fig.~\ref{Fig:QMEoS}, which include 46 EOSs of quark matter, i.e., 6 of them obtained with equivparticle model, 32 with perturbation model, and 8 with NJL model. Note that stiffer EOSs are obtained with larger $C$, $C_1$, $G_V$ and smaller $\Delta \mu$ in those models.

\section{\label{sec:MP} Mixed phase and quark-hadron interface}
\subsection{\label{sec:pasta} Inhomogeneous structures} 

If the surface tension of quark-hadron interface $\Sigma$ is smaller than the critical value $\Sigma_\mathrm{c}$, inhomogeneous structures of the mixed phase will emerge, i.e., the pasta phases. Adopting linearization for the charge densities of quark and hadron phases, the critical surface tension $\Sigma_\mathrm{c}$ can be estimated with~\cite{Voskresensky2003_NPA723-291}
\begin{equation}
\Sigma_\mathrm{c} = \frac{\left( \mu_{e0}^H - \mu_{e0}^Q \right)^2}{8 \pi \alpha \left(\lambda_D^Q + \lambda_D^H\right)},
\label{eq:sigma_c}
\end{equation}
where $\mu_{e0}^{H,Q}$ are the electron chemical potential and $\lambda_D^{H,Q}$ the Debye screening length of hadronic matter ($H$) and quark matter ($Q$) fulfilling both the $\beta$-stability condition and local charge neutrality condition. To obtain the properties of quark-hadron pasta phases, we adopt the formalism with a continuous dimensionality proposed by~\citet{Ravenhall1983_PRL50-2066}, where the energy density is determined by
\begin{equation}
E_\mathrm{t} = E_\mathrm{s} + E_\mathrm{C} + \chi E^\mathrm{I} + (1-\chi)E^\mathrm{II} \label{eq:Et}
\end{equation}
with
\begin{eqnarray}
E_\mathrm{s} &=& d\chi\frac{\Sigma}{r}, \\
E_\mathrm{C} &=& \frac{2\pi\alpha\chi r^2 {n_\mathrm{ch}^\mathrm{I}}^2}{(1-\chi)^2 (d+2)} \left[ \frac{2}{d-2} \left( 1-\frac{d}{2} \chi^{1-\frac{2}{d}} \right) +\chi \right]. \label{eq:E_c}
\end{eqnarray}
Here $\chi$, $r$, and $n_\mathrm{ch}^\mathrm{I}$ are the volume fraction, radius, and charge density of phase I, and $E^\mathrm{I, II}$ the corresponding energy densities. The continuous dimensionality $d$ lies in the range $1\leq d\leq3$, where $d=1$, 2, 3 represent the slab, rod/tube, droplet/bubble phases, respectively. For the case $d=2$, Eq.~(\ref{eq:E_c}) could yield the correct expression containing a logarithmic term~\cite{Wu2019_PRC99-065802}. The global charge neutrality condition $\chi n_\mathrm{ch}^\mathrm{I} + (1-\chi)n_\mathrm{ch}^\mathrm{II} = 0$ is fulfilled for the two phases in the cell. In Eq.~(\ref{eq:Et}), the term $E_\mathrm{s}$ represents the energy contribution of the quark-hadron interface, while $E_\mathrm{C}$ corresponds to the Coulomb energy per unit volume. Note that due to the charge screening effects, the local densities $n^\mathrm{I, II}$, $n_\mathrm{ch}^\mathrm{I, II}$, and $E^\mathrm{I, II}$ should in principal vary with space coordinates. For simplicity, we neglect such effects and the densities in each phase are assumed to be constants, which may affect our estimations on the sizes of the inhomogeneous structures at large $\Sigma$ (close to $\Sigma_\mathrm{c}$). However, for smaller $\Sigma$, the nonuniform distributions of charged particles in each phase become insignificant and assuming constant densities gives a fairly well description. In any cases, the negligence of charge screening effects has little impact on the obtained EOSs of MP.

The structures of MP can be fixed by minimizing the energy density in Eq.~(\ref{eq:Et}) at a given total baryon number density $n=\chi n^\mathrm{I} + (1-\chi)n^\mathrm{II}$. By taking derivatives of $E_\mathrm{t}$ with respect to each independent parameters ($r, \chi, d, n^\mathrm{I}, n_\mathrm{ch}^\mathrm{I}$) and equate them to zero, one obtains the following equations:
\begin{eqnarray}
&&{}E_\mathrm{s} = 2 E_\mathrm{C}, \\
&&{}P^\mathrm{I} - P^\mathrm{II} =
    \frac{d\Sigma\left( d\chi^{\frac{2}{d}} - d + \chi^{\frac{2}{d}-1}- \chi^{\frac{2}{d}}+1-\chi \right)}
         {r(1-\chi) \left(d\chi^{\frac{2}{d}}  - d - 2 \chi^{\frac{2}{d}} + 2 \chi^{\frac{2}{d}-1} \right)},\\
&&{}\frac{\left(d^3-12d+16\right)\chi-16}{\left( 2 d^2-8 \right) \ln(\chi) + d^3 - 12d}\chi^{\frac{2}{d}-1} =  1, \\
&&{}\mu_\mathrm{B}^\mathrm{I} = \mu_\mathrm{B}^\mathrm{II},  \\
&&{}\mu_{e}^\mathrm{I} - \mu_{e}^\mathrm{II} = \frac{d\Sigma}{r n_\mathrm{ch}^\mathrm{I}}.
\end{eqnarray}
Then the structures of MP are obtained by simultaneously fulfilling those equations, while the exact phase state (I = $H$ or $Q$) is determined for the case that gives a smaller $E_\mathrm{t}$. The quark fraction $\chi_Q$ is then fixed by
\begin{equation}
 \chi_Q =
 \left\{\begin{array}{c}
   \chi,  \ \ \ \  \text{I} = Q \\
   1-\chi, \text{I} = H \\
 \end{array}\right.. \label{Eq:chi}
\end{equation}
In practice, to further simplify our calculation, we expand the thermodynamic quantities with respect to $\mu_{e}$ at a given baryon chemical potential $\mu_\mathrm{B}$ as was done in Ref.~\cite{Xia2019_PRD99-103017}, where the chemical potential of each particle species is given by
\begin{equation}
\mu_i= B_i \mu_\mathrm{B} - q_i \mu_e.  \label{eq:weakequi}
\end{equation}
Here $B_i$ ($B_p=B_n=1$, $B_u=B_d=B_s=1/3$, and $B_e=B_\mu=0$) is the baryon number and $q_i$ ($q_p=1$, $q_n=0$, $q_u=2/3$, $q_d=q_s=-1/3$ and $q_e=q_\mu=-1$) the charge of particle type $i$. At given $\mu_\mathrm{B}$ and $\mu_e$, the pressure and energy densities are obtained with
\begin{eqnarray}
P &=& P_0 - \frac{1}{2} n_\mathrm{ch}' (\mu_e - \mu_{e0})^2,
\label{eq:P_lin} \\
E &=& E_0 + E' (\mu_e - \mu_{e0}) + \frac{1}{2} E''(\mu_e - \mu_{e0})^2. \label{eq:E_lin}
\end{eqnarray}
Here $P_0$, $E_0$, and $\mu_{e0}$ are the pressure, energy density, and electron chemical potential corresponding to those in Figs.~\ref{Fig:HMEoS} and~\ref{Fig:QMEoS}, while the derivatives $n_\mathrm{ch}' =\frac{\partial n_\mathrm{ch}}{\partial \mu_e},~~E' = \frac{\partial E}{\partial \mu_e},~~E'' = \frac{\partial^2 E}{\partial \mu_e^2}$ are taken at $\mu_e=\mu_{e0}$. The Debye screening length is related to $n_\mathrm{ch}'$ with $\lambda_\mathrm{D}\equiv \left( -4\pi \alpha n_\mathrm{ch}'\right)^{-1/2}$. According to the basic thermodynamic relations, the charge density and baryon number density are obtained with $n_\mathrm{ch} = n_\mathrm{ch}'(\mu_e - \mu_{e0})$ and $n = (E + \mu_e n_\mathrm{ch} + P)/\mu_\mathrm{B}$.

\subsection{\label{sec:the_surf}The quark-hadron interface}
At the quark-hadron interface, the wave functions of quarks approach to zero due to the presence of a confinement potential, where quarks are depleted and the corresponding energy contribution can be treated with a surface tension $\Sigma$. Based on MIT bag model~\cite{Oertel2008_PRD77-074015}, linear sigma model~\cite{Palhares2010_PRD82-125018,
Pinto2012_PRC86-025203, Kroff2015_PRD91-025017}, NJL model~\cite{Garcia2013_PRC88-025207, Ke2014_PRD89-074041}, three-flavor Polyakov-quark-meson model~\cite{Mintz2013_PRD87-036004}, Dyson-Schwinger equation approach~\cite{Gao2016_PRD94-094030}, equivparticle model~\cite{Xia2018_PRD98-034031}, nucleon-meson model~\cite{Fraga2019_PRD99-014046}, and Fermi gas approximations~\cite{Lugones2019_PRC99-035804, Lugones2017_PRC95-015804}, recent estimations suggest that the surface tension is likely small and $\Sigma \lesssim 30\ \mathrm{MeV/fm}^{2}$. Nevertheless, larger $\Sigma$ was also predicted in other investigations~\cite{Wen2010_PRC82-025809, Lugones2013_PRC88-045803, Alford2001_PRD64-074017}.

By counting the number of depleted quarks, the average effects due to quark depletion are treated with a modification to the density of states, i.e., the multiple reflection
expansion (MRE) method~\cite{Berger1987_PRC35-213, *Berger1991_PRC44-566, Madsen1993_PRL70-391, Madsen1993_PRD47-5156, Madsen1994_PRD50-3328}. Consider only the surface term, the modification for each quark flavor $i$ ($i=u,d,s$) reads~\cite{Berger1987_PRC35-213, *Berger1991_PRC44-566}
\begin{eqnarray}
\frac{\mbox{d} N_i^\mathrm{MRE}}{\mbox{d} p_i}  = - \frac{g_i p_i}{4\pi^2}\mathrm{arctan}\left(\frac{m_i}{p_i}\right) S, \label{eq:state_surf}
\end{eqnarray}
where $N_i^\mathrm{MRE}$ is the negative number of depleted quarks, $p_i$ the momentum of quarks, and $S$ the surface area. The corresponding contribution to the surface tension for each quark species $i$ is then obtained by equating the pressure $P_i^\mathrm{MRE}=-\Sigma_i^\mathrm{MRE}\frac{\mbox{d} S}{\mbox{d} V}$, which gives
\begin{eqnarray}
\Sigma_i^\mathrm{MRE} &=& \frac{1}{S}\int_0^{\nu_i} \frac{\mbox{d} N_i^\mathrm{MRE}}{\mbox{d} p_i} \left(\sqrt{p_i^2+m_i^2} - \sqrt{\nu_i^2+m_i^2}\right) \mbox{d} p_i \nonumber\\
                      &=& \frac{g_i m_i^3}{48\pi^2} \left[ (4x_i-3\pi) \sqrt{x_i^2+1}+2\pi - 2 \mathrm{arcsh}(x_i) \right. \nonumber\\
                      &&{}\left. + 2 (x_i^2+1)^{3/2}\mathrm{arccot}(x_i) \right].  \label{eq:sigmai}
\end{eqnarray}
Here $x_i\equiv \nu_i/m_i$ with $\nu_i$ being the Fermi momentum of quarks. The surface tension predicted by the MRE method is then obtained with $\Sigma = \Sigma^\mathrm{MRE} = \sum_{i=u,d,s} \Sigma_i^\mathrm{MRE}$.

Nevertheless, the MRE method tends to overestimate the surface tension by twice the value obtained with equivparticle model~\cite{Xia2018_PRD98-034031, Xia2019_AIPCP2127-020029}. This is mainly due to the different confinement potential adopted in those models, where the bag mechanism of the MRE method introduces an infinite wall that results in a sharp density discontinuity. Since the potential between quarks is proportional to the distance instead of a wall~\cite{Belyaev1984_PLB136-273}, a more realistic scenario was obtained with equivparticle model where confinement can be reached with density dependent quark masses in Eq.~(\ref{eq:mnbC}). A smoothly varying quark density is then obtained on the interface, where the surface tension was found to be connected with the density of quark matter by $\Sigma\approx 14.3 n^Q + 1.3$ (in $\mathrm{MeV/fm}^{2}$)~\cite{Xia2019_AIPCP2127-020029}. Meanwhile, it was shown that the surface tension predicted by the MRE method coincides with equivparticle model if we introduce a dampening factor, i.e.,  $\Sigma = 0.3\sum_{i=u,d,s} \Sigma_i^\mathrm{MRE}$. Note that the surface tension obtained by the equivparticle model~\cite{Xia2018_PRD98-034031, Xia2019_AIPCP2127-020029} was for the quark-vacuum interface. The contributions from the hadron phase can be roughly included by replacing the density of quark matter $n^Q$ by the density difference $\Delta n\equiv |n^Q - n^H|$ between the two phases. To avoid complications in minimizing the energy density in Eq.~(\ref{eq:Et}), we fix $\Sigma$ at a given total baryon number density $n$ for all cases considered here. Nevertheless, the surface tension estimated with the MRE method or equivparticle model varies with density, which will alter the baryon chemical potential and pressure with
\begin{eqnarray}
\mu_\mathrm{B} &=&  \mu_\mathrm{B}^\mathrm{I,II} + \frac{d\chi}{r} \frac{\mbox{d}\Sigma}{\mbox{d}n},\\
P &=& \mu_\mathrm{B} n - E_\mathrm{t}.
\end{eqnarray}
The corresponding EOSs of MP will thus become stiffer if $\frac{\mbox{d}\Sigma}{\mbox{d}n}>0$, which is the case in our current study.

Due to the uncertainties in $\Sigma$, in this work we adopt 9 different values, i.e.,
\begin{itemize}
  \item $\Sigma = 0$ with Gibbs construction;
  \item $\Sigma = 5$, 20, 50 $\mathrm{MeV/fm}^{2}$;
  \item $\Sigma = 0.5\Sigma_\mathrm{c}$ with $\Sigma_\mathrm{c}$ predicted by Eq.~(\ref{eq:sigma_c});
  \item $\Sigma = \Sigma^\mathrm{MRE}$ and $\Sigma = 0.3 \Sigma^\mathrm{MRE}$;
  \item $\Sigma = 14.3 \Delta n + 1.3$;
  \item $\Sigma > \Sigma_\mathrm{c}$ with Maxwell construction.
\end{itemize}
We have adopted both the Gibbs and Maxwell constructions in the two extreme scenarios with $\Sigma = 0$ and $\Sigma > \Sigma_\mathrm{c}$, a detailed description on those phase construction schemes can be found in our previous publication~\cite{Xia2019_PRD99-103017}.

\section{\label{sec:star}Results and discussions}

\begin{figure}
\includegraphics[width=\linewidth]{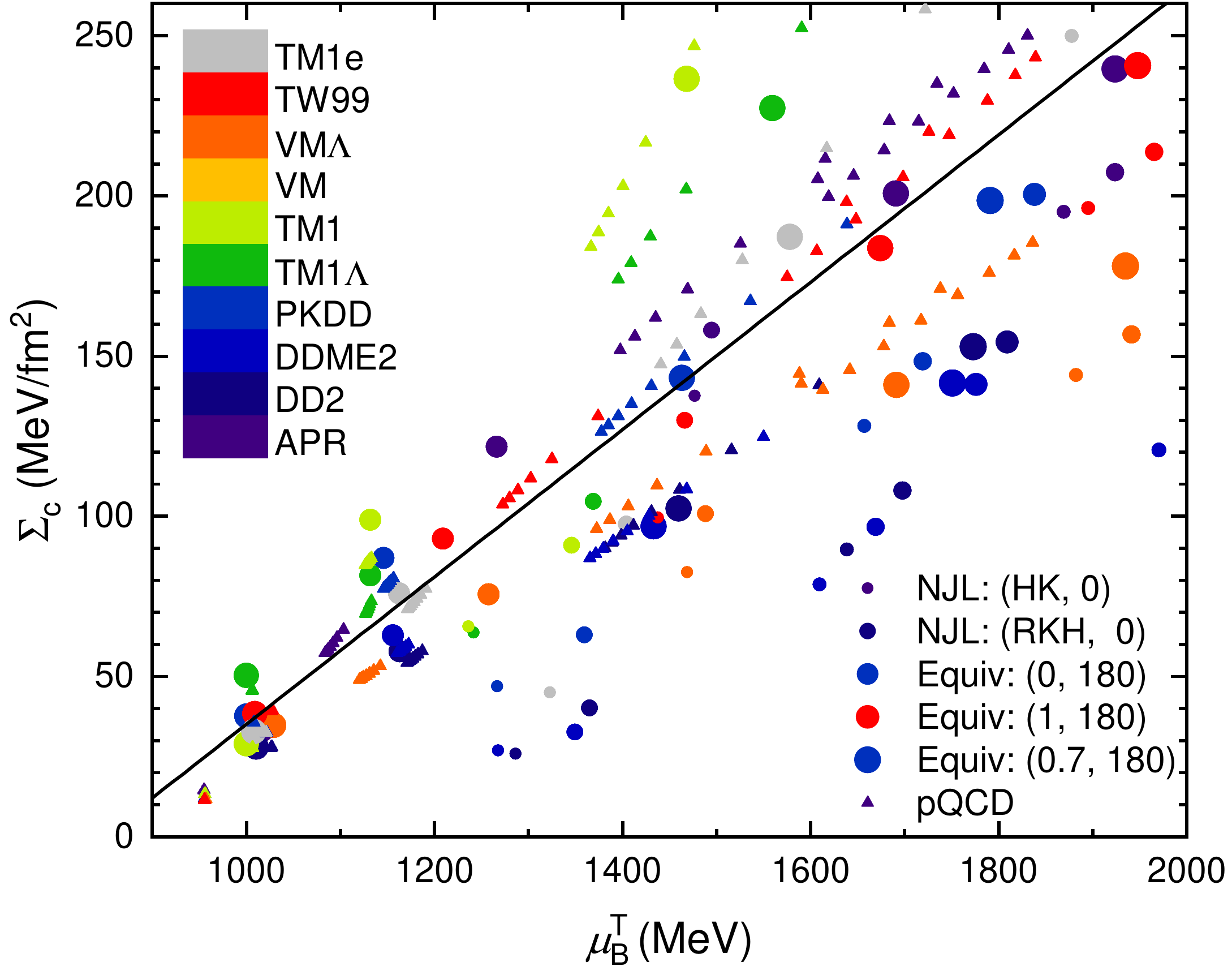}
\caption{\label{Fig:SigmaC_muT} The critical surface tension $\Sigma_\mathrm{c}$ estimated with Eq.~(\ref{eq:sigma_c}) as a function of the chemical potential $\mu_\mathrm{B}^\mathrm{T}$ on the occurrence of deconfinement phase transition. The symbol color indicates the hadronic EOSs adopted, while the size and shape represent the adopted quark model. }
\end{figure}

By equating the pressures of hadronic matter in Fig.~\ref{Fig:HMEoS} and quark matter in Fig.~\ref{Fig:QMEoS}, we obtain the critical chemical potential $\mu_\mathrm{B}^\mathrm{T}$ at which deconfinement phase transition occurs. In Fig.~\ref{Fig:SigmaC_muT} the corresponding critical surface tension $\Sigma_\mathrm{c}$ fixed by Eq.~(\ref{eq:sigma_c}) are presented. As the sizes of the full circles increase, the adopted model parameters for quark matter evolve in the order NJL: (HK, $G_V/G_S =$ 0$\rightarrow$1.5), NJL: (RKH,  $G_V/G_S =$ 0$\rightarrow$1.5), and Equiv with ($C$, $\sqrt{D}$ in MeV): (0, 180)$\rightarrow$(1, 140)$\rightarrow$(1, 180)$\rightarrow$($-0.2$, 180)$\rightarrow$(0.7, 140)$\rightarrow$(0.7, 180). In our previous study~\cite{Xia2019_PRD99-103017}, we have found a linear correlation $\Sigma_\mathrm{c} = 0.23(\mu_\mathrm{B}^\mathrm{T}-930)+19$ with $\Sigma_\mathrm{c}$ in MeV/fm${}^2$ and $\mu_\mathrm{B}^\mathrm{T}$ in MeV, which is indicated in Fig.~\ref{Fig:SigmaC_muT} with a black solid line. However, such a linear correlation fails to reproduce most of the current results in Fig.~\ref{Fig:SigmaC_muT}. In particular, we notice that the slope and intercept of the line vary with the adopted EOSs for both HM and QM. The inclusion of hyperons also plays a role if we adopt the effective interaction TM1 for the RMF model, while those obtained with cluster variational methods (VM and VM$\Lambda$) are not affected due to the much larger onset density of $\Lambda$-hyperons with the inclusion of three-baryon forces.

\begin{figure}
\includegraphics[width=\linewidth]{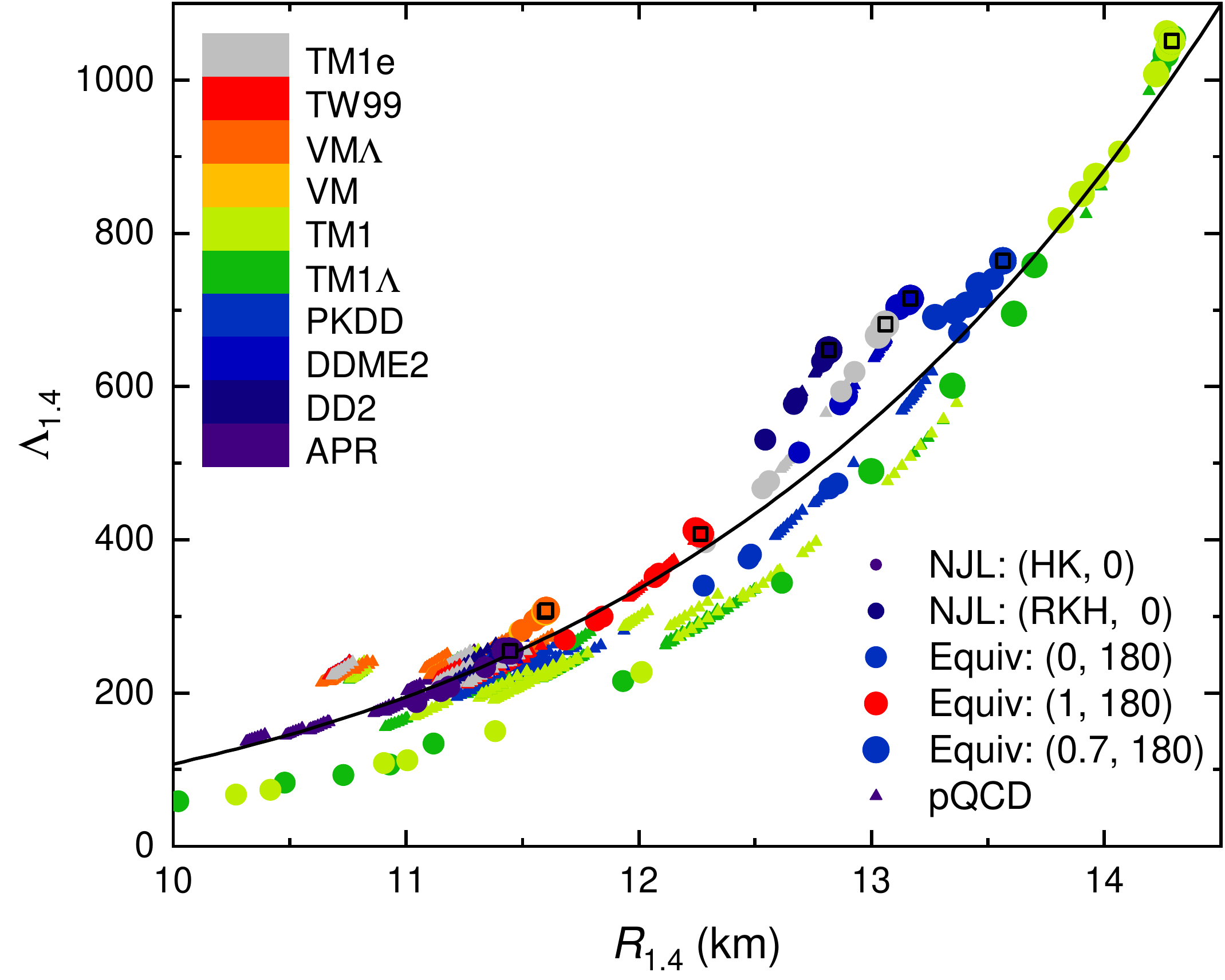}
\caption{\label{Fig:RLrelation} Correlation between tidal deformability and radius for $1.4 M_\odot$ compact stars. Same schemes as Fig.~\ref{Fig:SigmaC_muT} are adopted for the symbols, while the open squares correspond to the cases without a deconfinement phase transition.}
\end{figure}

With the properties of hadronic matter and quark matter determined in Sec.~\ref{sec:theHM} and Sec.~\ref{sec:theQM}, the structures of MP inside compact stars are obtained by minimizing the energy density in Eq.~(\ref{eq:Et}) with the surface tension $\Sigma$ fixed in Sec.~\ref{sec:the_surf}. This indicates in total 4084 EOSs, where the corresponding structures of hybrid stars are determined by solving the TOV equation~(\ref{eq:TOV}). Meanwhile, the tidal deformabilities of those stars are estimated with Eq.~(\ref{eq:td}). In Fig.~\ref{Fig:RLrelation} we present the obtained tidal deformability ($\Lambda_{1.4}$) as a function of radius ($R_{1.4}$) for $1.4 M_\odot$ compact stars, which shows a correlation between those two observable. In general, the traditional neutron stars indicated with open squares have the largest radius and tidal deformability, which will decrease as we include the quark phase. In most cases, for a given hadronic EOS, there are stronger correlations between $\Lambda_{1.4}$ and $R_{1.4}$ in hybrid stars, while the inclusion of hyperons has little impact on the relation. For traditional neutron stars, in Fig.~\ref{Fig:RLrelation} the black curve indicates the relation $\Lambda_{1.4}=5.9\times 10^{-5} R_{1.4}^{6.26}$ obtained in Ref.~\cite{Tsang2019_PLB796-1}, while the correlation between the maximum mass $M_\mathrm{max}$ and $\Lambda_{1.4}$ found in Ref.~\cite{Zhang2020_PRC101-034303} is not observed due to the first-order deconfinement phase transition.

\begin{figure}
\includegraphics[width=\linewidth]{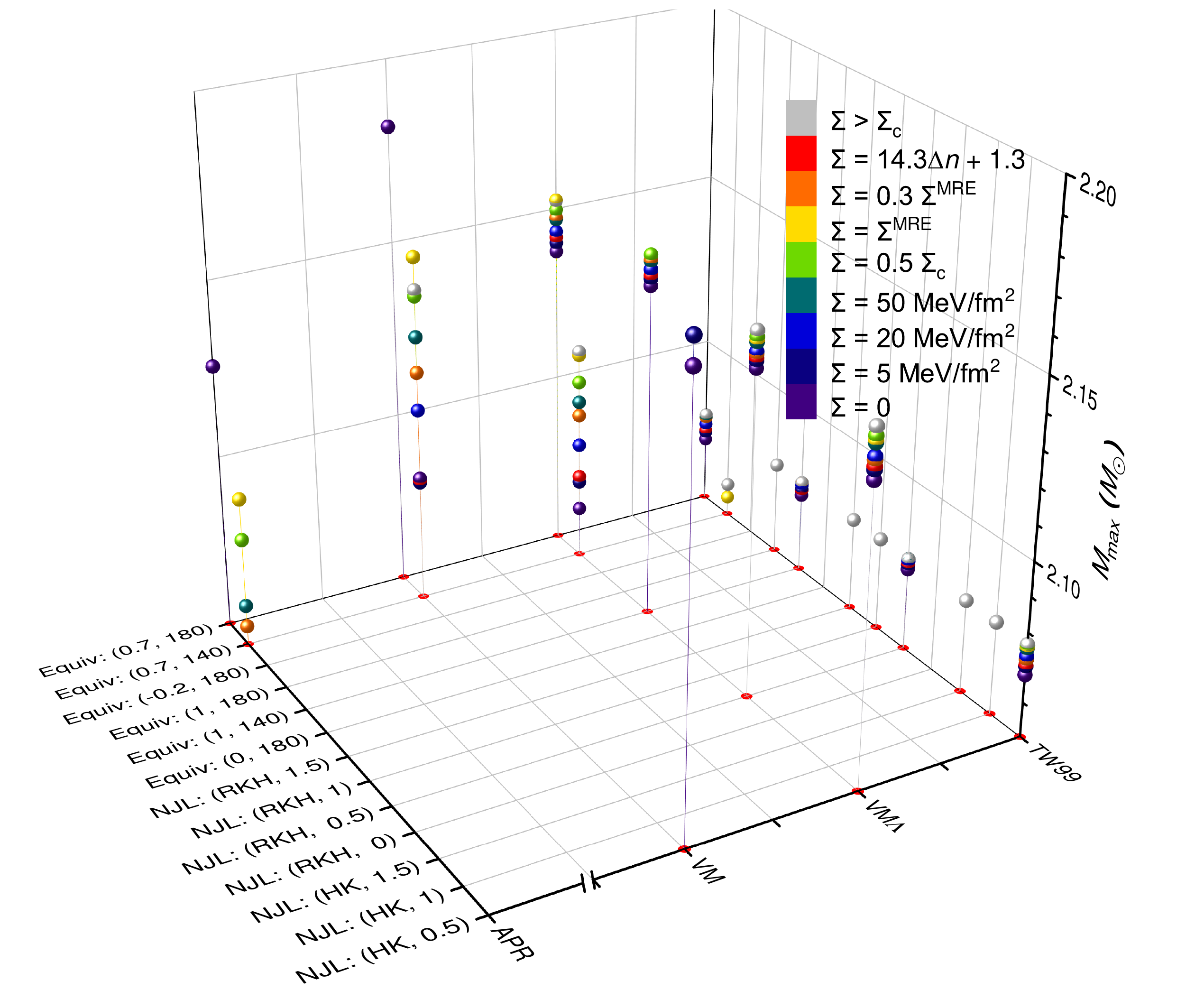}
\caption{\label{Fig:Mmax_NJLEquiv} The permitted maximum masses of hybrid stars predicted by various combinations of hadronic matter EOSs and quark matter EOSs. For NJL model, the parameter set (HK~\cite{Hatsuda1994_PR247-221} or RKH~\cite{Rehberg1996_PRC53-410}, $G_V/G_S$) is indicated explicitly. Similarly, the parameter set ($C$, $\sqrt{D}$ in MeV) for equivparticle model is presented as well. }
\end{figure}

\begin{figure*}
\includegraphics[width=0.5\linewidth]{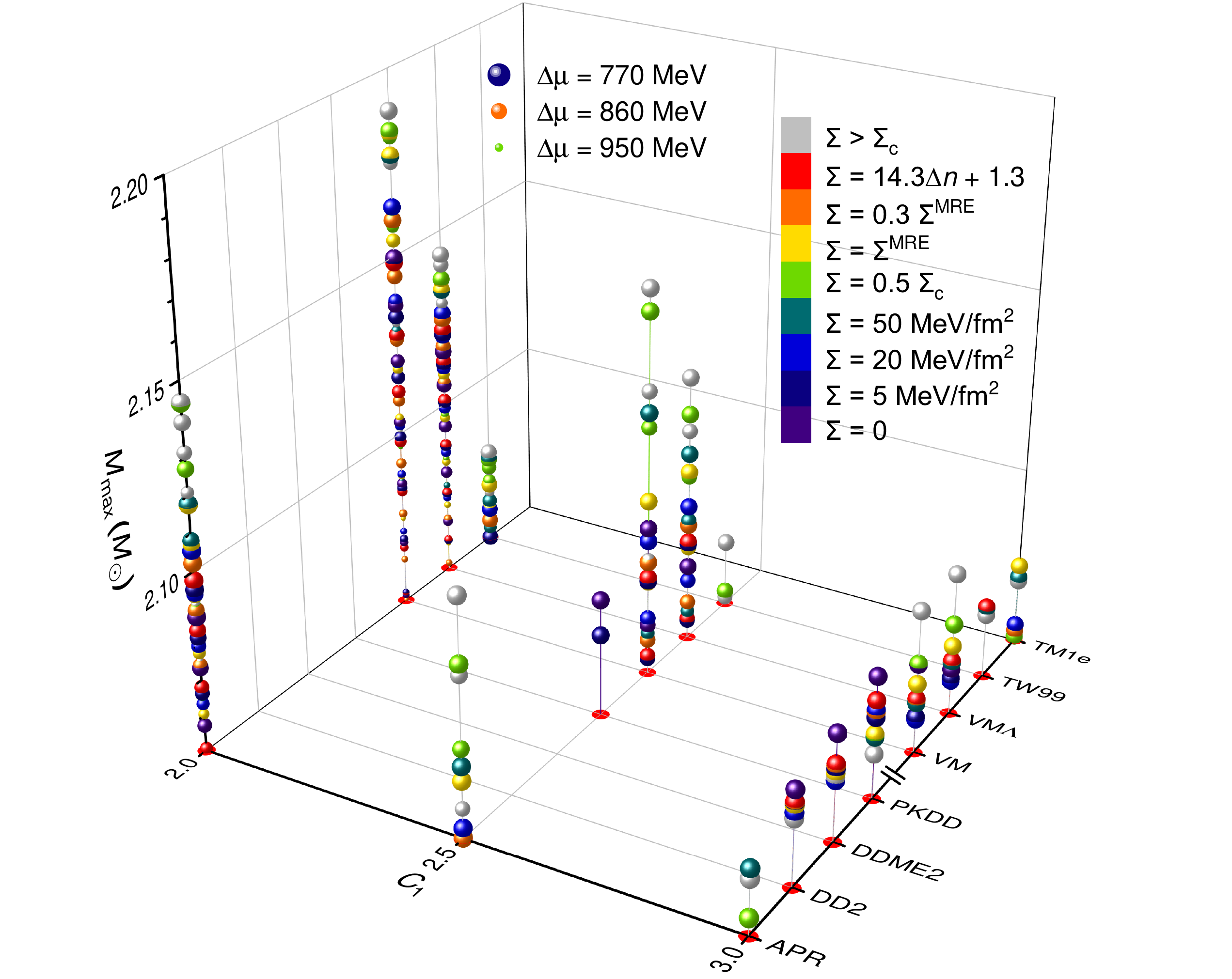}~
\includegraphics[width=0.5\linewidth]{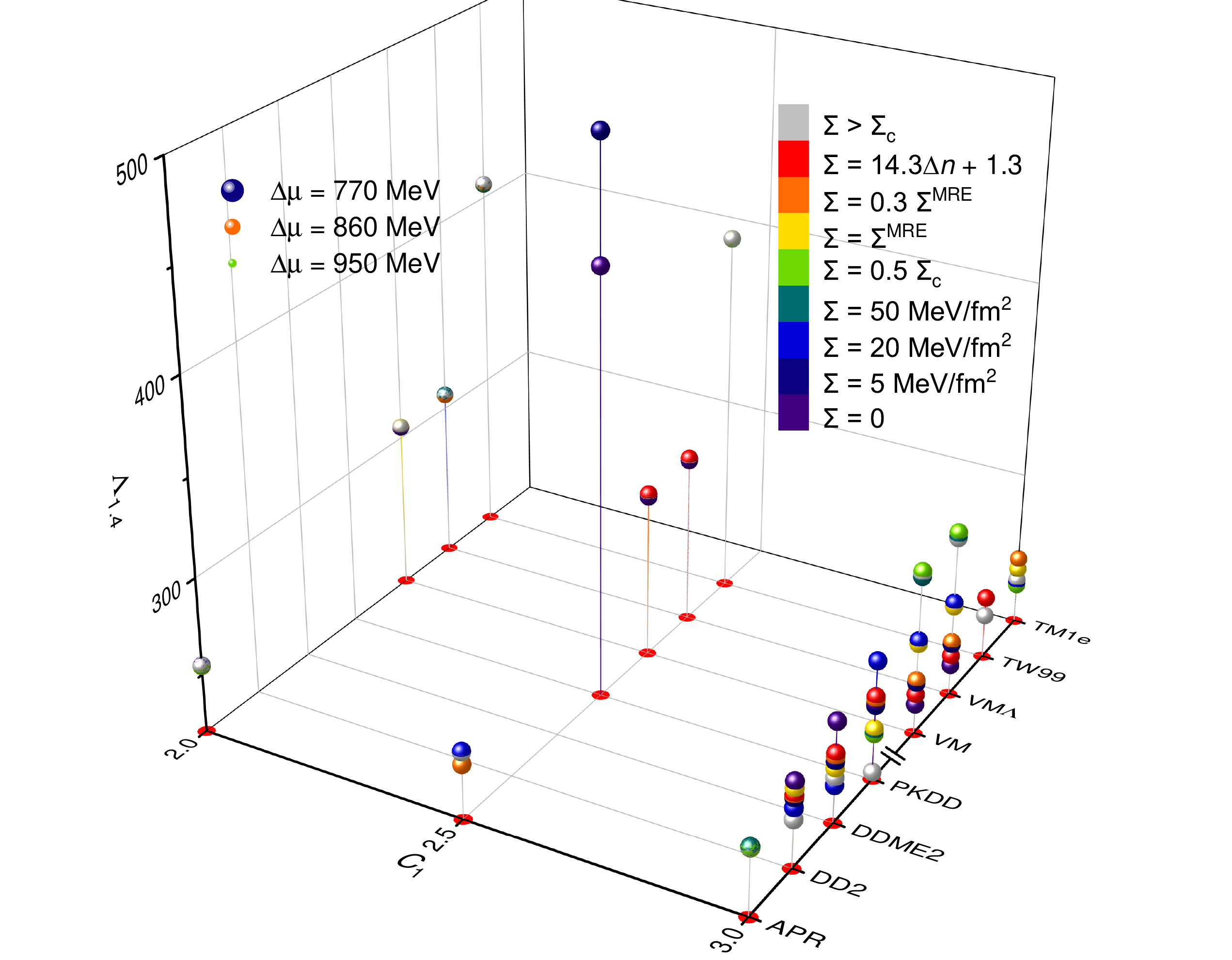}
\caption{\label{Fig:ML_pQCD} Same as Fig.~\ref{Fig:Mmax_NJLEquiv} but with the quark matter properties determined by perturbation model, where the maximum masses (left) and tidal deformabilities (right) of $1.4 M_\odot$ hybrid stars are presented. The model parameter $\Delta\mu$ is denoted by the size of each symbol.}
\end{figure*}

By comparing with the observational mass ($2.14^{+0.10}_{-0.09}M_{\odot}$) of PSR J0740+6620~\cite{Cromartie2020_NA4-72} as well as the tidal deformability constraint $70\leq \Lambda_{1.4}\leq 580$ from the GW170817 binary neutron star merger event~\cite{LVC2018_PRL121-161101}, we obtain the permitted combinations of hadronic and quark EOSs along with different values of surface tensions. In addition, we require the hadron-quark transition density $n^\mathrm{T}$ at $\mu_\mathrm{B}^\mathrm{T}$ is larger than 0.2 fm$^{-3}$ according to heavy-ion collision phenomenology, while the sound speed of hybrid star matter dose not exceed $c$. The permitted combinations of parameters are obtained, where in Figs.~\ref{Fig:Mmax_NJLEquiv} and \ref{Fig:ML_pQCD} we present the corresponding maximum masses and tidal deformabilities.

\begin{figure}
\includegraphics[width=\linewidth]{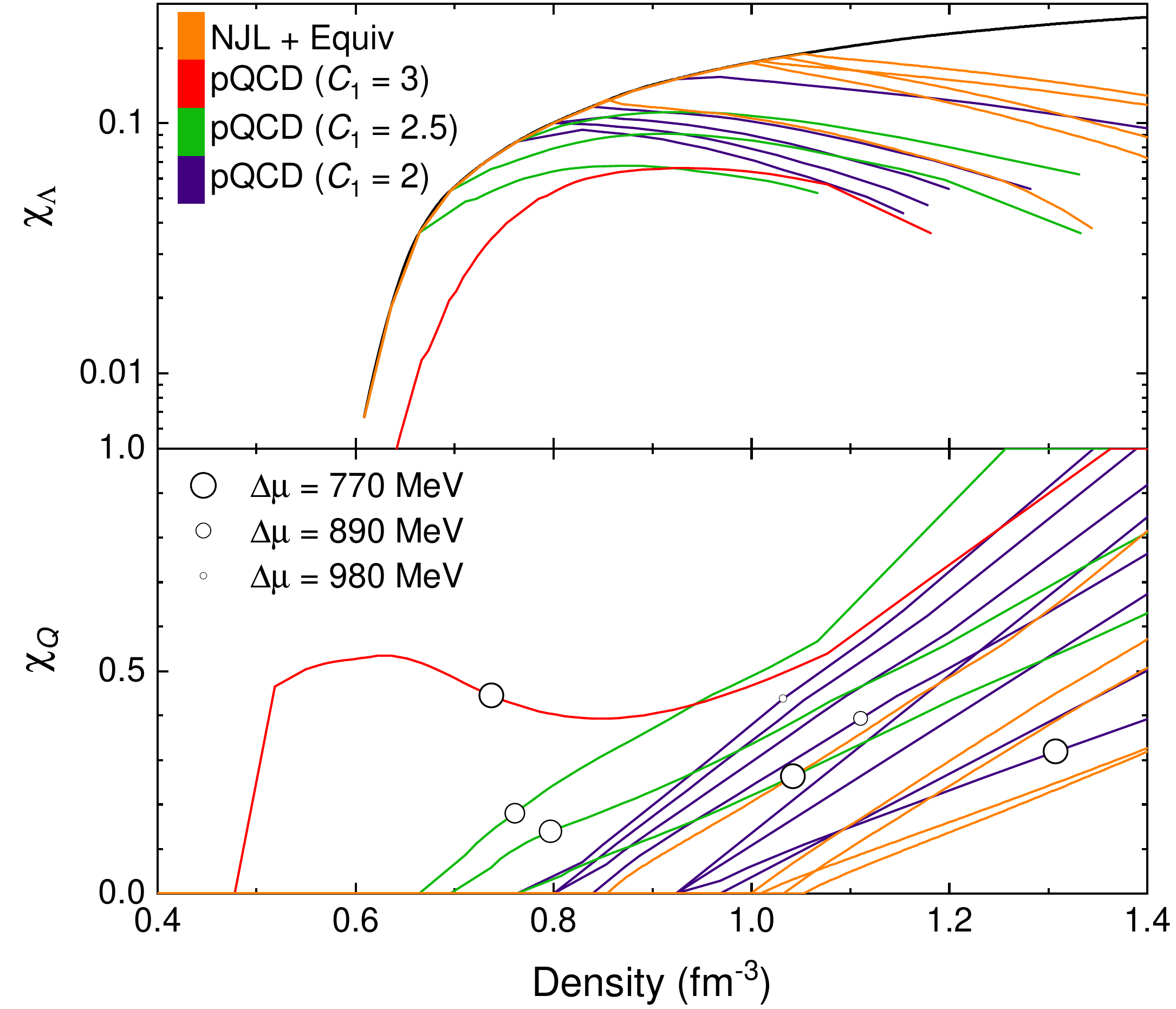}
\caption{\label{Fig:xLxQ} Hyperon ($\chi_\Lambda\equiv n_\Lambda/n$) and quark fractions ($\chi_Q$) of quark-hadron mixed phase in hybrid star matter as functions of baryon number density, which are obtained by adopting the hadronic EoS VM$\Lambda$ and $\Sigma = 50$ MeV/fm${}^2$. The black curve corresponds to the cases with $\chi_Q=0$.}
\end{figure}

\begin{figure}
\includegraphics[width=\linewidth]{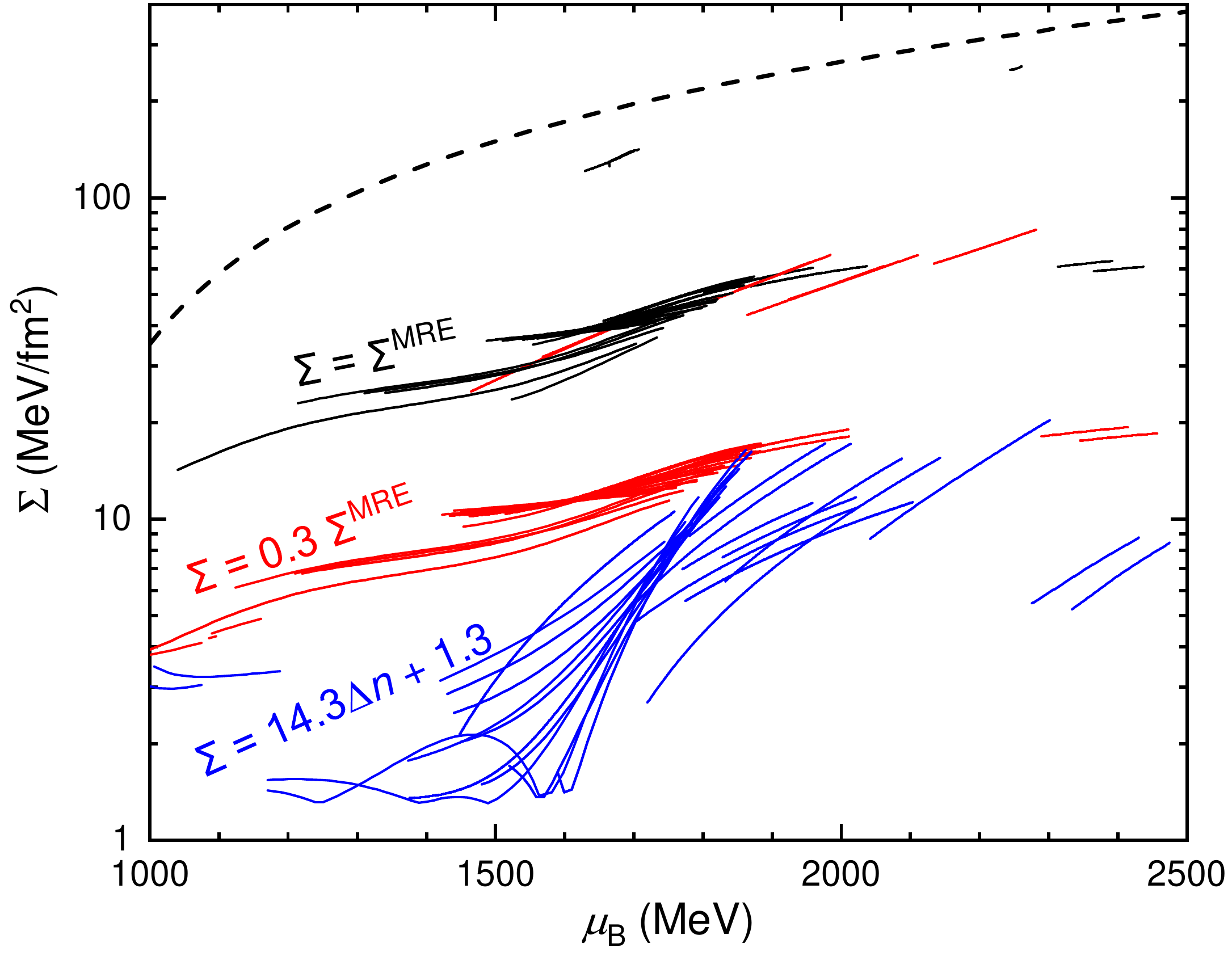}
\caption{\label{Fig:sigma_all} The surface tension $\Sigma$ obtained with various methods. The dashed curve corresponds to the critical surface tension obtained with $\Sigma_\mathrm{c} = 0.23(\mu_\mathrm{B}^\mathrm{T}-930)+19$ by taking $\mu_\mathrm{B}^\mathrm{T} = \mu_\mathrm{B}$~\cite{Xia2019_PRD99-103017}.}
\end{figure}

For the cases where quark matter properties are determined by equivparticle model and NJL model, the obtained onset density for quark matter usually exceeds the central density of $1.4 M_\odot$ neutron star if the two-solar-mass constraint is reached, in which case quark matter does not appear and $\Lambda_{1.4}$ coincides with those of traditional neutron stars. Nevertheless, quark matter persists in the most massive compact stars, where in Fig.~\ref{Fig:Mmax_NJLEquiv} we present the maximum masses of hybrid stars that are consistent with the aforementioned constraints. For the choices of hadronic EOSs, only APR, VM, VM$\Lambda$, and TW99 persist due to the constraint of the tidal deformability as indicated in Fig.~\ref{Fig:MLH}. For APR, VM, and VM$\Lambda$, the maximum mass is reduced since we require the the sound speed $v<c$. In most of the cases, the obtained $M_\mathrm{max}$ increases with $\Sigma$, while a slight deviation is observed for a few cases if $\Sigma$ is not constant, e.g., $\Sigma = \Sigma^\mathrm{MRE}$ or $\Sigma = 14.3 \Delta n + 1.3$. As indicated in Fig.~\ref{Fig:sigma_all}, the obtained $\Sigma$ increases with density, so that the corresponding EOSs of hybrid star matter are stiffer. In such cases, the maximum mass of hybrid stars may be even larger than those obtained with Maxwell construction at $\Sigma > \Sigma_\mathrm{c}$, e.g., the combinations of VM and equivparticle model. A comparison between the values of $M_\mathrm{max}$ obtained with VM and VM$\Lambda$ shows that the structures of the most massive compact stars are altered by the emergence of hyperons, despite the deconfinement phase transition occurred in the center of a compact star. This can be easily identified according to the hyperon and quark fractions indicated in Fig.~\ref{Fig:xLxQ}, where the $\Lambda$-hyperon appears before QM at $n\approx 0.6$ fm${}^{-3}$ and reaches a fraction of $\chi_\Lambda\approx 0.1$. The corresponding EOSs thus become softer in comparison with those of VM, where hyperons persist in MP. The main reason for this to occur is due to the fact that the energy per baryon of QM obtained with equivparticle model or NJL model is much larger than HM at $n\lesssim 0.8$ fm${}^{-3}$, so that $\chi_Q$ is reduced and the onset densities of QM are larger than hyperons.

\begin{figure*}
\includegraphics[width=0.7\linewidth]{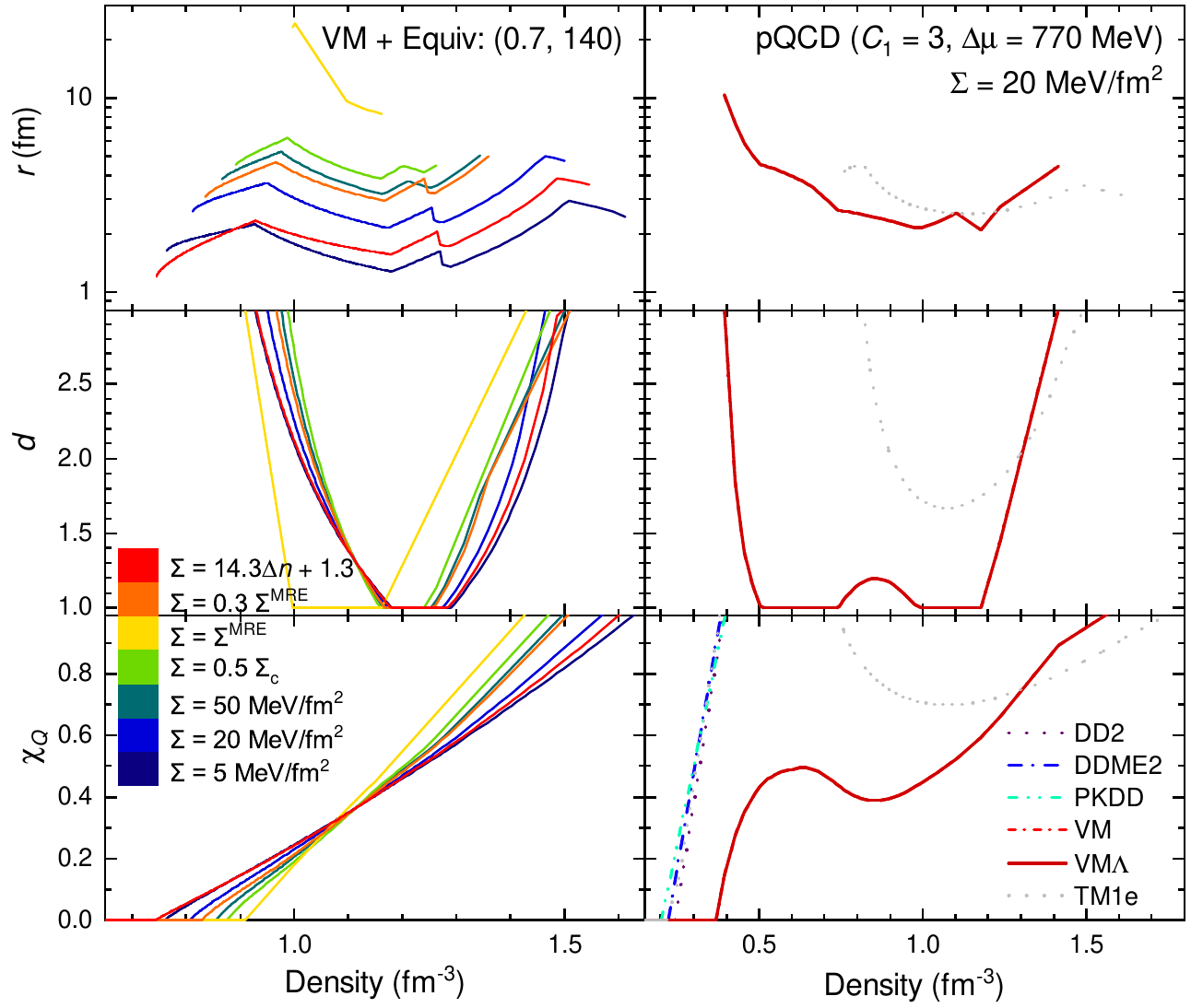}
\caption{\label{Fig:chid_all} The radius ($r$), dimensionality ($d$), and quark fraction ($\chi_Q$) of quark-hadron mixed phase in hybrid star matter as functions of baryon number density, which correspond to several representative cases obtained with various surface tension values (Left panel) and hadronic EOSs (Right panel).}
\end{figure*}

For those where the quark matter properties are obtained with perturbation model, the corresponding hadron-quark transition density $n^\mathrm{T}$ is small for large enough $C_1$. In such cases, beside the most massive ones, the structures of $1.4 M_\odot$ hybrid stars are also altered by the quark phase. This can be observed in Fig.~\ref{Fig:ML_pQCD}, where the maximum masses and tidal deformabilities of $1.4 M_\odot$ hybrid stars are presented. At $C_1 \gtrsim 2.5$, the obtained maximum mass $M_\mathrm{max}$ and tidal deformability $\Lambda_{1.4}$ usually increase with $\Sigma$, while there are few exceptions, e.g., PKDD, DDME2, and DD2. Note that in our previous study using TW99, sizable variation on the tidal deformability was observed with respect to $\Sigma$~\cite{Xia2019_PRD99-103017}, which is not evident in Fig.~\ref{Fig:ML_pQCD} since we have ruled out the cases with $M_\mathrm{max}<2.05\ M_\odot$. If the surface tension $\Sigma$ increases with density, the tidal deformability and maximum mass of hybrid stars may become larger. The obtained EOSs of hybrid star matter are softer for larger $\Delta\mu$, which predicts smaller $M_\mathrm{max}$. Nevertheless, it is worth mentioning that in most cases increasing $\Sigma$ will increase $M_\mathrm{max}$ more evidently in comparison with decreasing $\Delta\mu$. For the cases of VM and VM$\Lambda$, adopting small $C_1$ will give similar conclusion as in Fig.~\ref{Fig:Mmax_NJLEquiv} since the corresponding quark matter is too unstable to completely exclude hyperons, where the maximum mass is reduced with the emergence of hyperons. However, if we take $C_1 = 3$, as indicated in Fig.~\ref{Fig:xLxQ}, quark matter becomes more stable so that hyperons are suppressed with $\chi_\Lambda\lesssim 0.06$ due to a deconfinement phase transition. The structures of hybrid stars are thus hardly affected by hyperons. In any cases, hyperons do not appear in $1.4 M_\odot$ compact stars with the central density $n_\mathrm{central}\lesssim 0.57$ fm${}^{-3}$ so that $\Lambda_{1.4}$ are the same for those obtained with VM and VM$\Lambda$.

With the permitted combinations of hadronic and quark EOSs along with different values of surface tensions indicated in Figs.~\ref{Fig:Mmax_NJLEquiv} and \ref{Fig:ML_pQCD}, in Fig.~\ref{Fig:chid_all} we present the corresponding radius, dimensionality, and quark fraction of MP in hybrid star matter as functions of baryon number density. After the quark matter (phase I) appears and forms the droplet phase, the obtained dimensionality will later decrease from $d=3$ to $d=1$ as we increase the density. Then the phases I and II switch and the dimensionality increases from $d=1$ to 3. For the cases with dimensionality lies in between ($1<d<3$), the radius $r$ varies smoothly, while sudden variations are observed during the transition to $d=1$ or $3$. It is interesting to note that the structures of quark-hadron pasta phases vary smoothly by treating the dimensionality as a continuous variable with $d=1$--3, which may resemble the evolution of intermediate structures of droplet and rod, slab and tube found in the quantum molecular dynamics (QMD) simulations~\cite{Watanabe2003_PRC68-035806} as well as in the fully three-dimensional calculations adopting RMF model and Thomas-Fermi approximation~\cite{Okamoto2012_PLB713-284}. An early emergence of quark matter is observed if we adopt DD2, DDME2, PKDD, and TM1e for HM and perturbation model for QM, which is necessary in order to reduce the tidal deformabilities of $1.4 M_\odot$ hybrid stars that was otherwise too large for traditional neutron stars. In general, the quark fraction $\chi_Q$ increases with density, while there are several cases where $\chi_Q$ varies non-monotonically if we adopt the perturbation model for QM. At certain choices of parameters, a mixed phase may even appear after the formation of quark phase with $\chi_Q=1$, i.e., a retrograde transition with QM$\rightarrow$MP$\rightarrow$QM, which is mainly caused by adopting Eq.~(\ref{eq:BL}) for the quark phase. The transition with HM$\rightarrow$MP$\rightarrow$HM is also observed for few cases such as those obtained with NJL model, which is an artifact since the color confinement is not accounted for and we thus take $\chi_Q=0$, i.e., assuming a single phase of HM. The variation of $\chi_Q$ becomes more drastic if larger surface tension values were adopted and the density range of MP shrinks. Meanwhile, the obtained radii of phase I are usually on the order of fm and increase with $\Sigma$, which was discussed extensively in previous studies~\cite{Heiselberg1993_PRL70-1355, Voskresensky2002_PLB541-93, Tatsumi2003_NPA718-359, Voskresensky2003_NPA723-291, Endo2005_NPA749-333, Maruyama2007_PRD76-123015, Yasutake2014_PRC89-065803, Xia2019_PRD99-103017}.

In Fig.~\ref{Fig:sigma_all} we present the surface tension values estimated with various methods, which are much smaller than the critical surface tension $\Sigma_\mathrm{c}$ indicated in Fig.~\ref{Fig:SigmaC_muT}. This is a strong indication that the quark-hadron mixed phase prefers to form inhomogeneous structures inside hybrid stars, where the energy reduction $\Sigma_\mathrm{c}S$ arises from the relocation of charged particles is larger than the surface energy $\Sigma S$. Note that we have fixed the surface tension value at a given total baryon number density $n$, so that one does not need to worry about the variations of $\Sigma$ in minimizing the energy density in Eq.~(\ref{eq:Et}). In general, the surface tension values predicted by various methods are increasing with $\mu_\mathrm{B}$, where the formula $\Sigma = 14.3 \Delta n + 1.3$ gives the smallest $\Sigma$. The corresponding EOSs for MP are thus stiffer than those obtained with constant surface tension values, which increase the maximum masses of hybrid stars as indicated in Figs.~\ref{Fig:Mmax_NJLEquiv} and \ref{Fig:ML_pQCD}. Note that at $\mu_\mathrm{B}\approx 1400$ MeV, for few cases the perturbation model predicts smaller baryon number density of quark matter than that of hadronic matter, which causes fluctuations if the surface tension is estimated with $\Sigma = 14.3 \Delta n + 1.3$. At larger $\mu_\mathrm{B}$, the relation $0.3\Sigma^\mathrm{MRE} \approx 14.3 \Delta n + 1.3$ can be fulfilled approximately, which coincide with the predictions of equivparticle model~\cite{Xia2018_PRD98-034031, Xia2019_AIPCP2127-020029}. With the surface tension values indicated in Fig.~\ref{Fig:sigma_all}, the obtained structures of quark-hadron mixed phase follow the same trend as indicated in Fig.~\ref{Fig:chid_all}, where the radius $r$ increases with $\Sigma$.

\begin{figure}
\includegraphics[width=\linewidth]{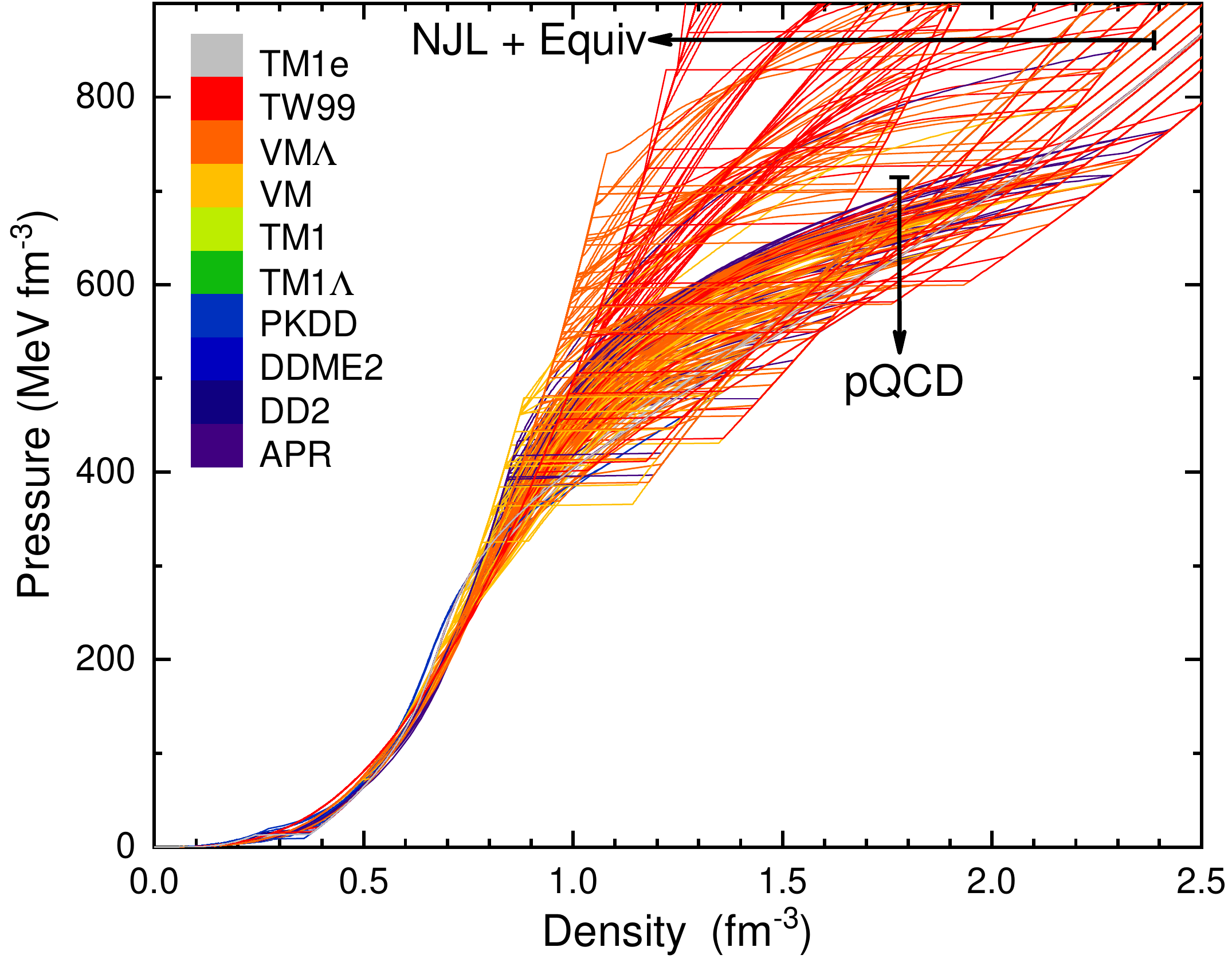}
\caption{\label{Fig:EOS_all} The pressure of hybrid star matter as functions of density, which correspond to the permitted cases indicated in Figs.~\ref{Fig:Mmax_NJLEquiv} and \ref{Fig:ML_pQCD}. The same constraints are applied to the following figures as well.}
\end{figure}

The EOSs of hybrid star matter are presented in Fig.~\ref{Fig:EOS_all}, where at $n\lesssim 0.8$ fm${}^{-3}$ the obtained pressures are close to each other for cases with and without the emergence of quark matter. At larger densities, the uncertainty grows and the pressure obtained with NJL model as well as few cases of equivparticle model are much larger than those of perturbation model. Nevertheless, we mention that the pressure obtained with perturbation model is expected to be reasonable at highest densities since a perturbation expansion with respect to $\alpha_\mathrm{s}$ becomes more reliable~\cite{Fraga2005_PRD71-105014}. The EOSs of quark-hadron mixed phase is sensitive to the surface tension value $\Sigma$, which become softer with smaller density range if $\Sigma$ is larger. The corresponding structures of hybrid stars are thus varying with $\Sigma$ as well, where the maximum mass and tidal deformation are altered as indicated in Figs.~\ref{Fig:Mmax_NJLEquiv} and \ref{Fig:ML_pQCD}.

\begin{figure}
\includegraphics[width=\linewidth]{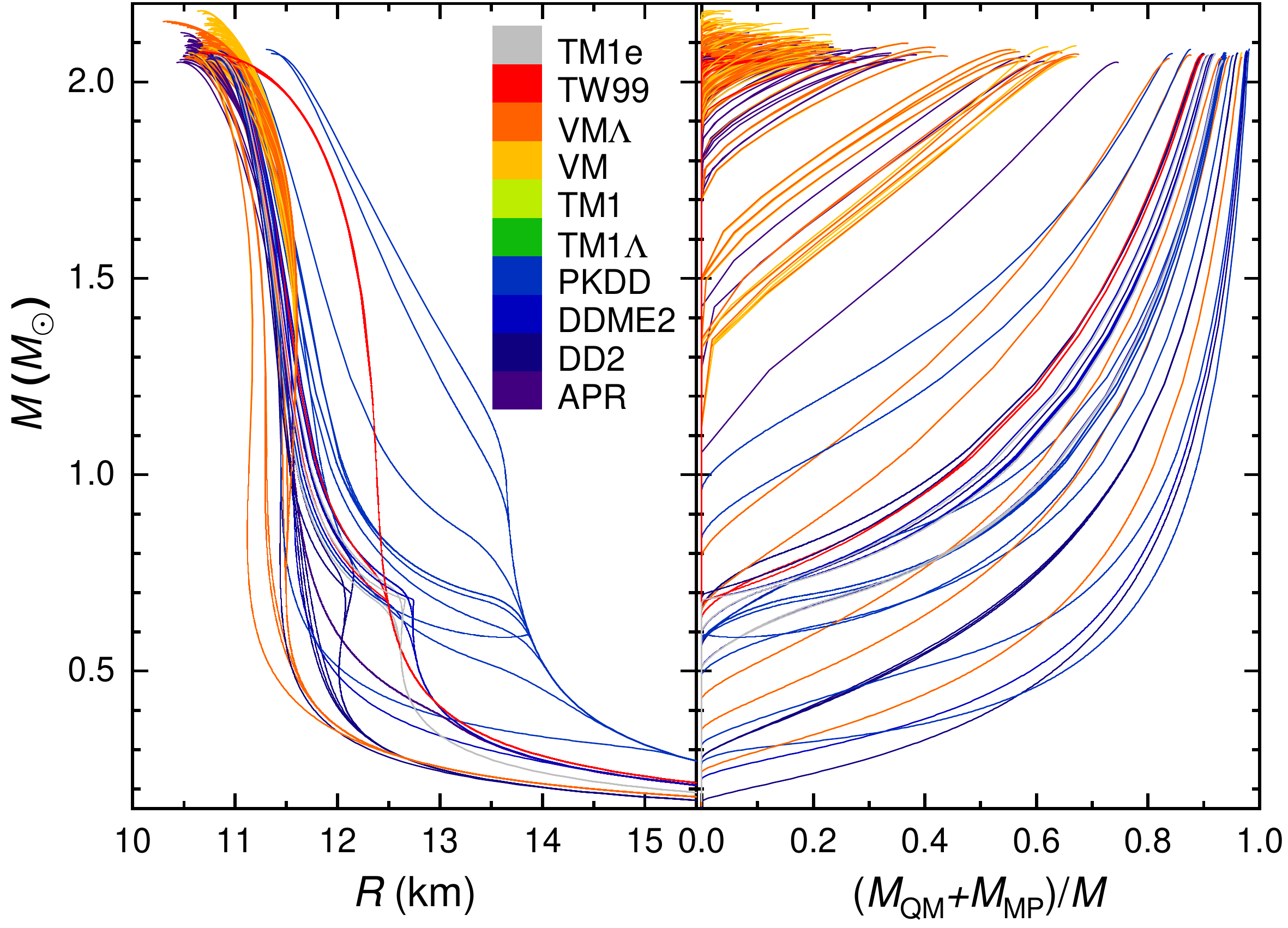}
\caption{\label{Fig:MRall} The mass-radius ($M$-$R$) relation and mass fraction of both QM and MP in hybrid stars obtained with the EOSs presented in Fig.~\ref{Fig:EOS_all}. }
\end{figure}

Based on the EOSs presented in Fig.~\ref{Fig:EOS_all}, the structures of compact stars can be obtained by solving the TOV equation~(\ref{eq:TOV}). In the left panel of Fig.~\ref{Fig:MRall} we present the corresponding $M$-$R$ relation of hybrid stars. The obtained radii of 1.4-solar-mass compact stars range from 11.1 km to 12.9 km, which are consistent with the radius measurements of J0030+0451~\cite{Riley2019_ApJ887-L21, Miller2019_ApJ887-L24} and binary neutron star merger event GW170817~\cite{LVC2018_PRL121-161101}. Meanwhile, the total mass of quark matter and quark-hadron mixed phase is obtained with $M_\mathrm{QM}+M_\mathrm{MP} = \int_0^{R_\mathrm{c}}4\pi E_\mathrm{t} r^2 \mbox{d}r$, where $R_\mathrm{c}$ is the critical radius that at $r>R_\mathrm{c}$ the quark fraction $\chi_Q$ reduces to 0. The corresponding fraction $(M_\mathrm{QM}+M_\mathrm{MP})/M$ is then indicated in the right panel of Fig.~\ref{Fig:MRall}. The obtained $M$-$R$ relation is identical to those in Fig.~\ref{Fig:MLH} before a deconfinement phase transition in the center. Once the quark phase emerges, the radius of a compact star becomes smaller and the compactness increases. For those with a smaller onset densities of quark phase (e.g., TM1e/PKDD/DDME2/DD2 \& perturbation model), the fraction $(M_\mathrm{QM}+M_\mathrm{MP})/M$ increases quickly starting at $M\approx 0.2\text{--}1\ M_{\odot}$ and approaches to almost 1 at $M=M_\mathrm{max}$. A third family of compact stars~\cite{Gerlach1968_PR172-1325, Benic2015_AA577-A40} is observed for the case with a combination of PKDD \& perturbation model ($C_1=3$, $\Delta\mu=770$ MeV) \& $\Sigma > \Sigma_\mathrm{c}$, where a jump of 152 MeV fm${}^{-3}$ in the energy density from nuclear matter to quark matter is predicted. Note that if we take a larger $\Delta\mu$ for Eq.~(\ref{eq:BL}), the third family of compact stars indicated in Fig.~\ref{Fig:MRall} will not be permitted by astrophysical observations. In almost all combinations of hadronic and quark EOSs along with different values of surface tensions, the quark phase persists in the most massive compact stars, where the fraction $(M_\mathrm{QM}+M_\mathrm{MP})/M$ ranges from 0 to almost 1. This is in coincidence with the recent studies~\cite{Annala2020_NP, Blaschke2020_Universe6-81, Blaschke2020_EPJA56-124}, which suggest the presence of quark-matter cores inside massive compact stars.

\begin{figure}
\includegraphics[width=\linewidth]{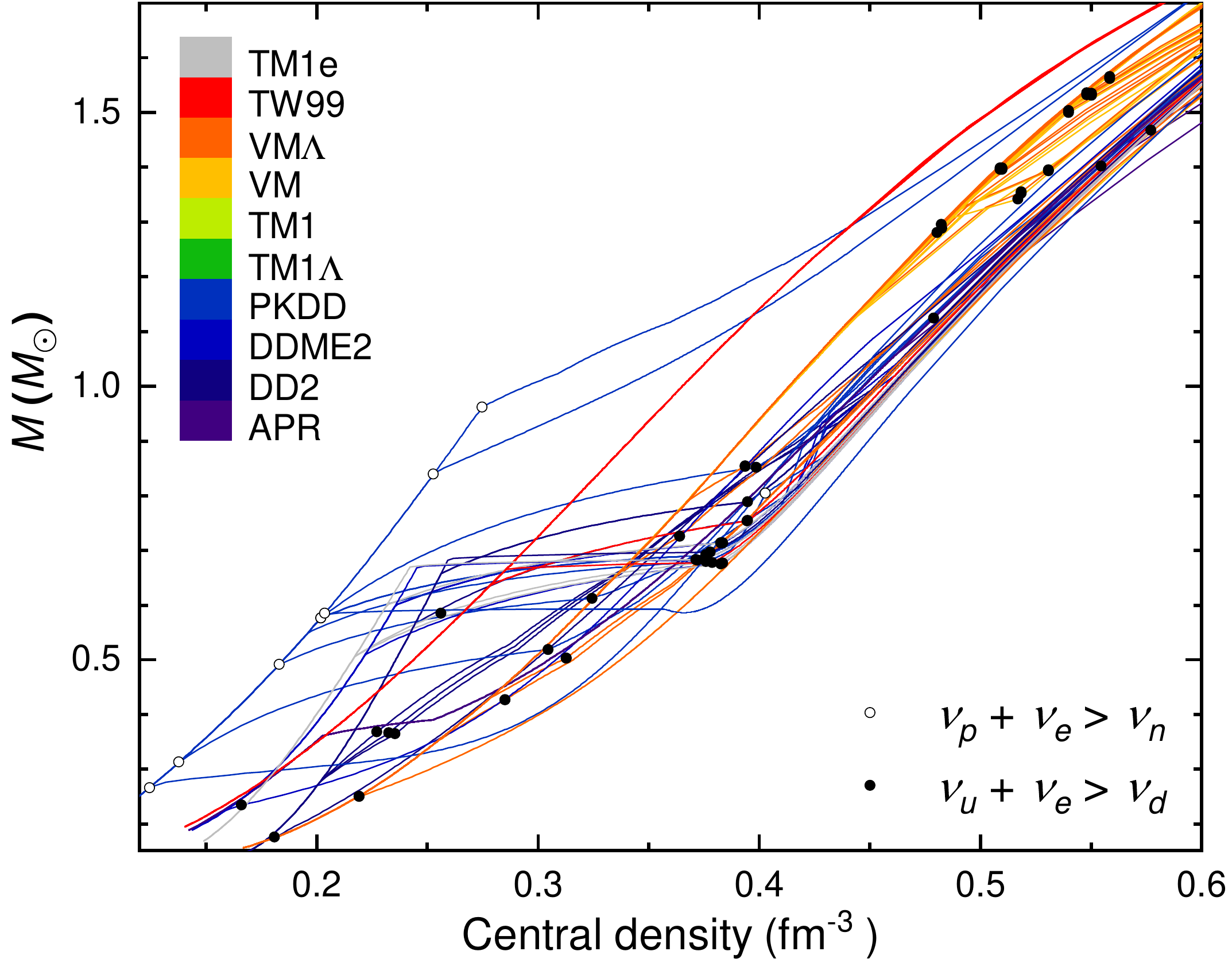}
\caption{\label{Fig:Mnall} Hybrid star masses as functions of the central density corresponding to those indcated in Fig.~\ref{Fig:MRall}. The open and full circles represent the critical mass and central density beyond which the direct Urca processes of rapid cooling may take place.}
\end{figure}

Aside from the observational constraints on the mass, radius, and tidal deformability, the thermal evolution of compact stars also provides important information on their internal composition~\cite{Page2006_NPA777-497}. Based on the thermal emission, kinematic measurements, spin period and its derivative, both the surface temperatures and ages of compact stars can be estimated~\cite{Vigano2013_MNRAS434-123, Potekhin2015_SSR191-171}. According to various observational data, the theoretical cooling models suggest that rapid cooling due to the direct Urca (DU) processes should not occur in typical neutron stars within the mass range 1--$1.5\ M_{\odot}$~\cite{Blaschke2004_AA424-979, Popov2006_AA448-327, Klaehn2006_PRC74-035802}. The DU process in nuclear matter involves the $\beta$-decay and electron capture processes of nucleons, i.e., $n\rightarrow p + e^-+\bar{\upnu}_e$ and $p + e^-\rightarrow n+\upnu_e$. The quark analogs of the nucleon DU processes are $d\rightarrow u + e^-+\bar{\upnu}_e$ and $u + e^-\rightarrow d+\upnu_e$. Those processes will occur inside compact stars once the momentum conservation is fulfilled, i.e., the triangle inequalities $\nu_n\leq \nu_p + \nu_e$ and $\nu_d\leq \nu_u + \nu_e$ with $\nu_i$ being the Fermi momentum~\cite{Pethick1992_RMP64-1133}. If the strangeness is involved, the DU processes such as $\Lambda\rightarrow p + e^-+\bar{\upnu}_e$ and $s\rightarrow u + e^-+\bar{\upnu}_e$ should also take effects. However, we neglect those processes here since their neutrino emissivities are expected to be less than that of nucleon/quark DU processes~\cite{Page2006_NPA777-497} while hyperons appear at rather large densities as indicated in Fig.~\ref{Fig:xLxQ}. In Fig.~\ref{Fig:Mnall} the hybrid star masses as functions of the central density are presented, where the open and full circles mark the critical mass ($M_\mathrm{DU}$) and central density ($n_\mathrm{DU}$) fulfilling the triangle inequalities. For stars with masses larger than $M_\mathrm{DU}$, it was shown that the neutrino emissivity is enhanced significantly by the DU processes~\cite{Spinella2018_Universe4-64}, which cool the stars too rapidly within just a few years~\cite{Blaschke2004_AA424-979}.

\begin{figure}
\includegraphics[width=\linewidth]{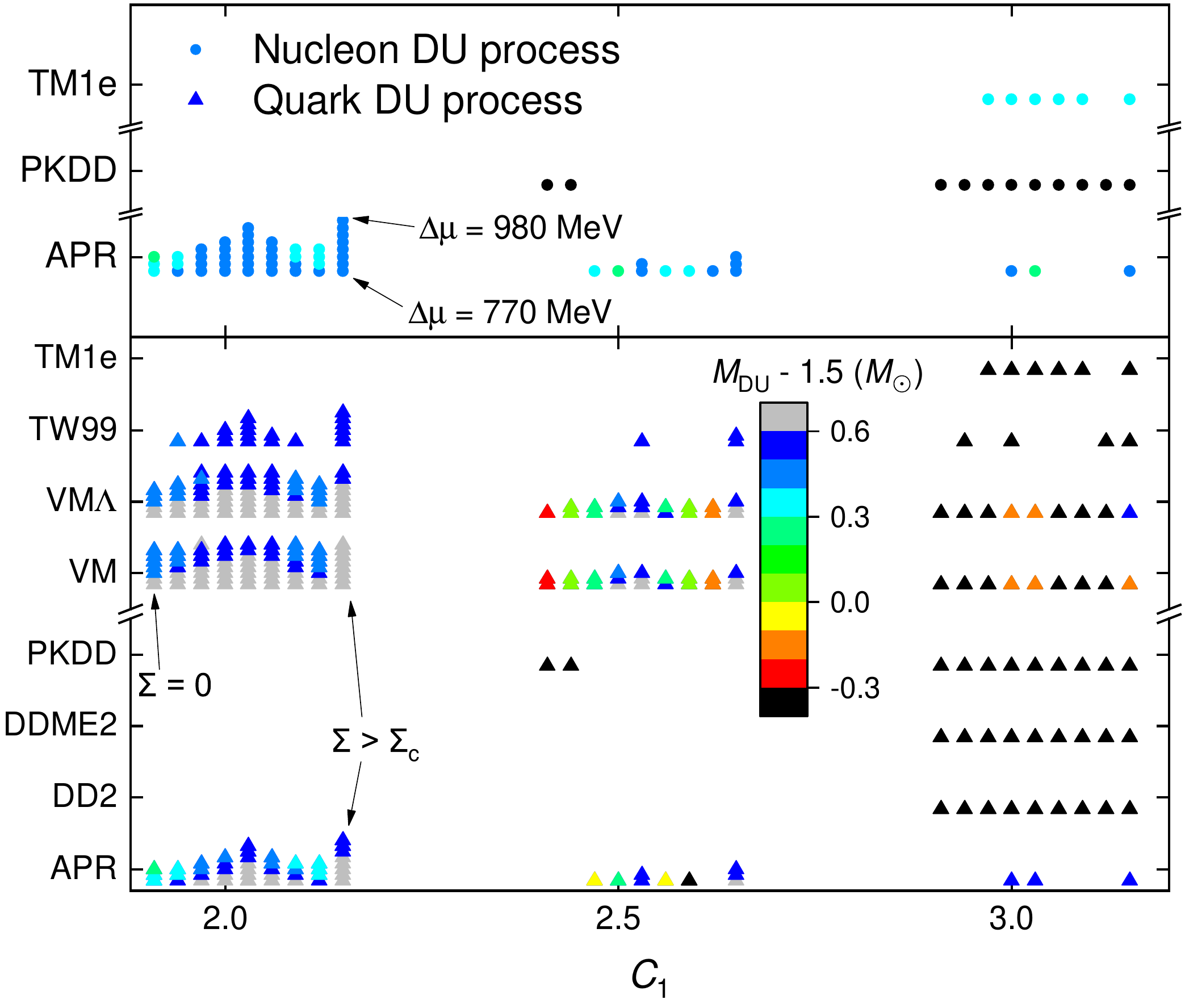}
\caption{\label{Fig:DUcrit} The critical mass $M_\mathrm{DU}$ substracted by $1.5 M_{\odot}$. For each combination of $C_1$ and hadronic EOS, the adopted surface tension value varies in the order $\Sigma = 0$, 5, 20, 50 $\mathrm{MeV/fm}^{2}$, $0.5\Sigma_\mathrm{c}$, $\Sigma^\mathrm{MRE}$, $0.3 \Sigma^\mathrm{MRE}$, $14.3 \Delta n + 1.3$, and $\Sigma > \Sigma_\mathrm{c}$ from left to right, while $\Delta\mu$ increases from bottom to top.}
\end{figure}

The obtained critical mass and central density corresponding to NJL model and equivparticle model are usually large with $M_\mathrm{DU}\gtrsim 2 M_{\odot}$ and $n_\mathrm{DU} \gtrsim 0.8$ fm${}^{-3}$, which are consistent the observational thermal evolution of compact stars~\cite{Blaschke2004_AA424-979, Popov2006_AA448-327, Klaehn2006_PRC74-035802}. If a large $C_1$ is adopted for the perturbation model, QM will emerge at small densities and lead to the quark DU processes. Meanwhile, as indicated in Fig.~\ref{Fig:Mnall}, the nucleon DU processes will take place if we adopt PKDD for HM, where the triangle inequality is fulfilled due to the reduction of electron chemical potentials with the appearance of QM. A more detailed investigation on different combinations of hadronic EOSs and surface tensions is indicated in Fig.~\ref{Fig:DUcrit}, where the critical masses $M_\mathrm{DU}$ for both nucleon and quark DU processes are presented. For these where DU processes never take place, we take $M_\mathrm{DU} = M_\mathrm{max}$. By applying the constraint $M_\mathrm{DU}>1.5 M_{\odot}$, it is found that most combinations of hadronic EOSs with quark EOSs determined by perturbation model at $C_1 = 3$ are not permitted due to an early emergence of QM at small hadron-quark transition densities $n^\mathrm{T}\lesssim 0.3\ \mathrm{fm}^{-3}$, which lead to the quark DU processes. Meanwhile, larger values of $n^\mathrm{T}$ are obtained with the hadronic EOSs VM, VM$\Lambda$ and APR, and consequently the quark DU processes do not occur if large surface tension values are adopted. In such cases, an early emergence of QM at $n^\mathrm{T}\lesssim 0.3\ \mathrm{fm}^{-3}$ is prohibited by the DU criterion. At $C_1 = 2.5$, the small surface tension values $\Sigma = 0$ and $14.3 \Delta n + 1.3$ are excluded for the hadronic EOSs VM and VM$\Lambda$ due to the occurrence of quark DU processes. Similarly, $\Sigma = 20$ $\mathrm{MeV/fm}^{2}$, $\Sigma^\mathrm{MRE}$, and $0.3 \Sigma^\mathrm{MRE}$ are not permitted for the hadronic EOS APR. As indicated in the upper panel of Fig.~\ref{Fig:DUcrit}, the hadronic EOS PKDD is excluded since nucleon DU processes always occur in typical compact stars, which rules out the third family of compact stars in Fig.~\ref{Fig:MRall}. It is worth mentioning that the color superconductivity of quark matter will effectively hinder the quark DU processes~\cite{Page2006_NPA777-497}, so that the cases with $M_\mathrm{DU}<1.5 M_{\odot}$ in the lower panel of Fig.~\ref{Fig:DUcrit} may not necessarily lead to a fast cooling and the tension with the observational data can be eased. For example, if QM forms a two-flavor superconducting phase, the cooling history of a hybrid star with a large quark core may be consistent with the X-ray data~\cite{Blaschke2000_ApJ533-406}. Note that in the extreme scenario where hybrid stars are comprised almost entirely of QM in the color-flavor-locked phase, heat capacity would be too low to be consistent with observations~\cite{Cumming2017_PRC95-025806, Horowitz2019_AP411-167992}.

\section{\label{sec:con}Conclusion}

In this work we investigate systematically the possible hadron-quark deconfinement phase transition in dense stellar matter, and its influence on compact star structures. For the hadronic phase, we adopt in total 10 different EOSs, i.e., 8 nuclear EOSs (TM1e~\cite{Shen2020_ApJ891-148}, TM1~\cite{Sugahara1994_NPA579-557}, PKDD~\cite{Long2004_PRC69-034319}, TW99~\cite{Typel1999_NPA656-331}, DDME2~\cite{Lalazissis2005_PRC71-024312}, DD2~\cite{Typel2010_PRC81-015803}, VM~\cite{Togashi2017_NPA961-78}, APR~\cite{Akmal1998_PRC58-1804}) and 2 hyperonic EOSs (TM1$\Lambda$~\cite{Sun2018_CPC42-25101} and VM$\Lambda$~\cite{Togashi2016_PRC93-035808}), which are predicted by relativistic-mean-field model~\cite{Meng2016_RDFNS} and variational method with realistic baryon interactions~\cite{Akmal1998_PRC58-1804, Togashi2017_NPA961-78}. For the quark phase, we adopt 46 EOSs predicted by equivparticle model~\cite{Peng2000_PRC62-025801, Wen2005_PRC72-015204, Xia2014_PRD89-105027}, perturbation model~\cite{Freedman1977_PRD16-1169, Fraga2005_PRD71-105014, Kurkela2010_PRD81-105021}, and NJL model with vector interactions~\cite{Hatsuda1994_PR247-221, Rehberg1996_PRC53-410}. With the properties of both hadronic matter and quark matter fixed, the structures of quark-hadron mixed phase are obtained assuming a continuous dimensionality as proposed by~\citet{Ravenhall1983_PRL50-2066}. The energy contribution due to the quark-hadron interface is treated with a surface tension $\Sigma$, where we have taken constant values for $\Sigma$ as well as those estimated by the multiple reflection expansion method~\cite{Berger1987_PRC35-213, *Berger1991_PRC44-566, Madsen1993_PRL70-391, Madsen1993_PRD47-5156, Madsen1994_PRD50-3328} and equivparticle model including both linear confinement and leading-order perturbative interactions~\cite{Xia2018_PRD98-034031, Xia2019_AIPCP2127-020029}. The critical surface tension $\Sigma_\mathrm{c}$ that accounts for the energy reduction due to the relocation of charged particles is estimated for various combinations of quark and hadronic EOSs. It is found that in most cases we have $\Sigma<\Sigma_\mathrm{c}$, where inhomogeneous structures for the quark-hadron mixed phase are favored.

As we increase the density of hadronic matter, quark matter will emerge and forms a quark-hadron mixed phase. By minimizing the energy density at given baryon number density, we have obtained the radius, dimensionality, and quark fraction of MP. It is found that the obtained radius normally ranges from $\sim$1 fm to $\sim$10 fm, and is increasing with $\Sigma$. The radius evolves more smoothly with density if the dimensionality changes continuously. Adopting various combinations of hadronic and quark EOSs along with different values of surface tensions, the quark fraction usually increases monotonically and turns into a pure quark phase. The corresponding EOSs for hybrid star matter are obtained, which predict the structures of compact stars by solving the TOV equation. It is found that the correlation between radius and tidal deformability in traditional neutron stars~\cite{Tsang2019_PLB796-1, Zhang2020_PRC101-034303} preserves in hybrid stars. Once quark matter emerges inside compact stars, the quark-hadron interface plays an important role on their structures. The surface tension $\Sigma$ estimated with the multiple reflection expansion method or equivparticle model increases with density, which predicts stiffer EOSs for the quark-hadron mixed phase and increases the maximum mass of hybrid stars. The hyperons are suppressed if we adopt a quark model that predicts relatively small energy per baryon of quark matter at small densities.  Based on various constraints of nuclear physics, causality limit, and pulsar observations, we obtain the permitted parameter sets that are consistent with observation. It is found that the quark phase persists inside the most massive compact stars in almost all the permitted cases. Meanwhile, comparing with higher density regions, the variation of pressure is small at $n\lesssim 0.8$ fm${}^{-3}$ irrespective of the emergence of quark matter. The current constraints can be further improved based on the thermal evolution of compact stars, which rules out an early emergence of quark matter at densities smaller than 0.3 $\mathrm{fm}^{-3}$ in the absence of color superconductivity.

\begin{acknowledgements}
This work was supported by National Natural Science Foundation of China (Grants No.~11705163, No.~11875052, No.~11525524, No.~11675083, and No.~11775119), JSPS KAKENHI (Grants No.~20K03951 and No.~20H04742), and Ningbo Natural Science Foundation (Grant No.~2019A610066). The support provided by China Scholarship Council during a visit of C.-J. X. to JAEA is acknowledged. The computation for this work was supported in part by the HPC Cluster of SKLTP/ITP-CAS and the Supercomputing Center, CNIC, of the CAS.
\end{acknowledgements}


\newpage

%

\end{document}